\documentclass[aps,prx,reprint,superscriptaddress]{revtex4-2-scholarone}
\usepackage{dcolumn}
\usepackage[dvipsnames]{xcolor}
\usepackage{bm}
\usepackage[utf8]{inputenc}
\usepackage[english]{babel}
\usepackage[T1]{fontenc}
\usepackage{tikz}
\usepackage[compat=0.6]{yquant}
\usepackage{soul} 
\usepackage{graphicx} 
\usepackage{algpseudocode}
\usepackage{amsmath}
\usepackage{amsfonts}
\usepackage{subcaption}
\usepackage{pgfplots}
\pgfplotsset{compat=1.18}
\usepackage{ulem}
\usepackage{hyperref}
\usepackage{wrapfig}
\usepackage{xparse}
\usepackage{comment}
\usepackage{framed}
\usepackage{placeins}
\usepackage{physics,tensor,siunitx}
\usepackage{booktabs}
\usepackage{multirow}
\usepackage{graphicx}
\usepackage{pgfplots}
\usepgfplotslibrary{groupplots}
\pgfplotsset{compat=1.18}

\usepackage{tikz}
\usepackage{pgfplots}

\usetikzlibrary{backgrounds}

\pgfdeclarelayer{background}
\pgfsetlayers{background,main}

\newcommand{\sn}[1]{\text{\scriptsize#1}}
\newcommand{\dket}[1]{\fcolorbox{black}{gray!20}{$\ket{#1}$}}

\newcommand{\threshold}{\ensuremath{\mathcal{T}}}

\definecolor{cbBlue}{HTML}{0072B2}
\definecolor{cbRed}{HTML}{E41A1C}
\definecolor{cbOrange}{HTML}{D55E00}
\definecolor{cbGreen}{HTML}{009E73}
\definecolor{cbPurple}{HTML}{CC79A7}
\definecolor{cbSky}{HTML}{56B4E9}
\definecolor{cbGray}{HTML}{999999}
\definecolor{shadecolor}{named}{yellow}

\pgfplotsset{
    compat=1.18,
    every axis/.append style={
        width=\columnwidth,
        height=0.68\columnwidth,
        tick label style={font=\small},
        label style={font=\small},
        legend style={
            font=\scriptsize,
            draw=none,
            fill=none,
            cells={anchor=west}
        },
        major grid style={cbGray!25},
        ymajorgrids=true,
        xmajorgrids=false,
        line width=0.9pt
    }
}
  
\begin{document}
\title{Half the Interference, Most of the Answer: \\
Approximate Quantum Simulation via Path-Sum Pruning}

\author{Sinan Pehlivanoglu \footnote{Corresponding Author}}%
 \email[Corresponding author: ]{spehliva@iu.edu}
\affiliation{%
 Department of Computer Science, Luddy School of Informatics, Computing, and Engineering, Indiana University, Bloomington, Indiana, USA
}
\affiliation{Indiana University Quantum Science and Engineering Center, Bloomington, Indiana 47405, USA}

\author{Srinivasan Iyengar}%
 \email{iyengar@iu.edu}
\affiliation{%
 Department of Chemistry, Department of Physics, Indiana University, Bloomington, Indiana, USA
}
\affiliation{Indiana University Quantum Science and Engineering Center, Bloomington, Indiana 47405, USA}

\author{Amr Sabry}%
 \email{sabry@iu.edu}
\affiliation{%
 Department of Computer Science, Luddy School of Informatics, Computing, and Engineering, Indiana University, Bloomington, Indiana, USA
}
\affiliation{Indiana University Quantum Science and Engineering Center, Bloomington, Indiana 47405, USA}

\date{\today}

\begin{abstract} 
Classical simulation of quantum circuits is expensive for two distinct reasons. The obvious one is state-space size: an $n$-qubit system requires exponentially many amplitudes. The less obvious one is interference: useful output distributions emerge only after many computational histories have been coherently combined at common endpoints, and this aggregation step is itself a substantial source of cost. We introduce \emph{statistical interference sampling}, a framework that makes this second bottleneck explicit by treating endpoint interference as a separately schedulable computation. Using the Chemical Abstract Machine (ChAM) as our model, weighted path contributions evolve as concurrent molecular species, and interference reactions combine contributions that share a common output state. A threshold rule terminates the process once an endpoint accumulates sufficient amplitude, discarding the remaining reactions.
The method does not improve worst-case complexity and is not intended as a general-purpose simulator. Its purpose is to ask a more targeted question: how much of the interference calculation can be skipped while still recovering a useful output distribution? Across repeated independent executions at each threshold, we
observe a circuit-dependent tradeoff among endpoint-interference
work, task-level success, and distributional fidelity. We use
Deutsch--Jozsa as a worked illustration and evaluate the method on
Grover search, Simon's problem, and Shor period-finding. We find that thresholding omits substantial
endpoint-interference work while attaining high observed task success. An early-termination policy can attain 100\% observed task success on
several Grover, Simon, and Shor configurations while omitting up to 65\% of the interference work. Circuits with strong amplitude separation, where a large gap exists between the amplitude accumulated at the correct endpoint and the maximum amplitude attainable at any incorrect endpoint, benefit most from thresholding; circuits with flatter output distributions, where valid outputs share comparable amplitudes across many endpoints rather than concentrating probability on a single marked state, are less amenable to this strategy.
These results suggest that interference arithmetic is a structured resource that admits meaningful approximation, and that exposing it explicitly opens new opportunities for pruning strategies across path-sum, Pauli-path, and tensor-network simulation methods. 
\end{abstract}
\maketitle
\section{Introduction}
\label{sec:intro}

Classical simulation of quantum circuits is usually described as hard
because of Hilbert-space dimension: an $n$-qubit pure state requires
$2^n$ complex amplitudes, and manipulating them scales exponentially.
This framing, while correct, obscures a second and more physically
transparent source of cost. Quantum algorithms do not merely create
large amplitude vectors; they rely on the coherent combination of
exponentially many computational histories at common measurement
endpoints. The useful output distribution emerges only after those
histories have interfered, reinforcing at some endpoints and canceling
at others. Treating this aggregation as a first-class computational
step, rather than an implicit consequence of matrix-vector
multiplication, is the central idea of this paper.

The path-integral formulation of quantum mechanics, due to
Feynman~\cite{FEYNPI}, makes the structure of this aggregation
precise: the amplitude for any transition is a coherent sum over all
computational histories connecting the initial and final states, with
each history contributing a complex weight. This representation is
useful here because it makes a structural property of an important
class of algorithms visible. Algorithms whose power
derives from amplitude amplification, including Grover
search and the quantum Fourier transform at the core
of Shor’s algorithm, are deliberately designed to concen-
trate constructive interference on a small set of endpoints
while destructive interference dominates everywhere else.
Most aggregation work in these circuits produces cancel-
lations that contribute nothing to the final distribution.
This asymmetry is the key property we exploit in our approach: pruning some paths early is likely to discard pre-
dominantly destructive contributions, leaving the output
distribution largely intact. As Feynman’s path-integral
formulation makes clear, the set of paths that interfere
constructively is generally far smaller than the set that
interferes destructively, and it is this imbalance that un-
derlies the emergence of discrete quantum states.
To act on this intuition we need a computational model in which
interference reactions are explicit and interruptible. We use the
ChAM~\cite{Berry1992}, a rule-based model of concurrent computation
in which a collection of weighted entities evolves by local reactions.
In our setting, each entity represents a partial path contribution,
carrying a basis state, a circuit-depth label, and a complex amplitude.
Reaction rules then implement the three steps of the simulation: time
evolution, which advances each contribution through the circuit;
endpoint tagging, which marks contributions that have reached the final
depth; and amplitude aggregation, which combines tagged contributions
that share a common output state. Because ChAM execution is
asynchronous, these reactions can occur in any order and the simulation
can be stopped at any point without corrupting the partial results
already accumulated. We model this interruption by introducing a threshold parameter~$\threshold$. Once the set of active states contains a state $\alpha\dket{b}$ such that the state is ready for interference and satisfies $|\alpha|\geq\threshold$, the
simulator halts, discards unfinished reactions,
forms one complex aggregate per endpoint from the tracked sampling
population, and samples after normalizing the squared aggregate
magnitudes. A high threshold recovers the full
saturated simulation whereas a low threshold discards more reactions and risks more error.

Statistical interference sampling differs from Pauli-path
truncation and tensor-network bond truncation in the computational
stage it targets. While existing methods reduce the set
of contributions that must be propagated; statistical interference sampling reduces how completely the surviving contributions must be aggregated. These
approaches may therefore be complementary.

The empirical question is how this tradeoff behaves for
circuits with different endpoint-amplitude profiles. We evaluate the
implementation on standard benchmark circuits and use selected IBM
Brisbane histograms as qualitative hardware context. The results show a
circuit- and threshold-dependent tradeoff among task success,
distributional fidelity, and endpoint-interference work. We find that nearly 65\% of endpoint interference reactions can be omitted while maintaining over
90\% output accuracy for most algorithms tested. Our main
contributions are as follows.

\begin{itemize} 

\item We recast circuit simulation as a discrete path-sum problem in which endpoint interference is an explicit aggregation operation, separable from time evolution. 

\item We give a ChAM semantics for this aggregation, with distinct reaction families for time evolution, endpoint tagging, and amplitude combination. 

\item We define a threshold-based termination rule that stops the interference process before saturation and quantify the resulting approximation error.

\item We use Deutsch--Jozsa to illustrate the framework and
evaluate it on Grover search, Simon's problem, and Shor
period-finding instances. Across 50 repeated fresh-JVM trials, we report
task-level success with Wilson confidence intervals, direct
measurements of endpoint-interference reduction, total-variation
distance with bootstrap uncertainty, and complementary Hellinger
metrics. We compare selected output histograms qualitatively with IBM
Brisbane data.

\end{itemize} 

The remainder of the paper is organized as follows. Section~\ref{sec:ChAM} formalizes the path-sum representation and the ChAM interference model, establishing the three reaction families that govern simulation. Section~\ref{sec:threshold} defines the threshold-based termination rule and the resulting partial-distribution sampling procedure. Section~\ref{sec:example} works through the Deutsch-Jozsa circuit as a concrete illustration of the simulation mechanics and thresholding behavior. Section~\ref{sec:background} introduces the benchmark algorithms and describes the experimental setup. Section~\ref{sec:empirical} presents the empirical evaluation, including qualitative comparisons with selected IBM Brisbane histograms. Section~\ref{sec:related} situates the method relative to path-sum, Pauli-path, and tensor-network simulation approaches. Section~\ref{sec:conc} summarizes limitations and future directions. Appendix~\ref{sec:formal} provides the formal probabilistic analysis, including a finite-population concentration bound for threshold-crossing errors under an idealized random-arrival model. Appendix~\ref{app:hellinger} reports complementary Hellinger-distance results, while Appendix~\ref{app:tensor_network_connection} develops the connection to tensor-network contraction and bond truncation. The implementation is publicly available at \url{https://github.com/sinanspd/partial-quantum-interference-effects}.

\section{Path-Sum Representation and ChAM Semantics}
\label{sec:ChAM}

A quantum circuit of depth $D$, defined as the number of gate layers, maps an initial basis state $\ket{x_0}$
to a superposition over output states. In the path-sum picture, this
evolution decomposes into a sum over all computational histories
connecting $\ket{x_0}$ to each possible output. A history, or path,
is a sequence of intermediate basis states
\[
    \mu=(x_0,x_1,\ldots,x_D),
\]
and each path contributes a complex weight
\[
    w(\mu)=\prod_{i=1}^{D}\bra{x_i}U_i\ket{x_{i-1}}
\]
determined by the gates along the way. The amplitude for reaching
output $x_D$ is then the coherent sum of all path weights that end
there:
\[
    A_D(x_D)=\sum_{\mu:\,x_D} w(\mu).
\]
This is the discrete analogue of Feynman's path integral~\cite{FEYNPI},
and it makes the structure of quantum measurement explicit. The
observable probability of outcome $x_D$~is
\[
    P(x_D) = |A_D(x_D)|^2 =
    \left|\sum_{\mu:\,x_D} w(\mu)\right|^2 =
    \sum_{\mu,\mu':\,x_D} w(\mu)\,w(\mu')^*
\]
a double sum over pairs of paths sharing the same endpoint. The
cross-terms $w(\mu)w(\mu')^*$ with $\mu \neq \mu'$ are precisely the
interference contributions: they can reinforce or cancel depending on
the relative phases of the two path weights. Interference is therefore
not a global property of the state vector but a local operation that
aggregates pairs of paths at common endpoints.

This observation motivates the ChAM model~\cite{Berry1992}. Rather
than computing the full state vector implicitly through matrix-vector
multiplication, we represent each path contribution as a molecule in a
chemical solution and implement endpoint aggregation as an explicit
reaction. A molecule $\alpha\ket{b}^{d_i}$
carries a basis state $b$, a circuit-depth label $d_i$, and a complex
amplitude $\alpha$ summarizing the coherent evolution of the path up
to depth $d_i$. The simulation proceeds through three reaction
families.

\textsc{Time evolution} applies the next gate and branches each
molecule into its successors:
\[
    \alpha\ket{b}^{d_i}
    \;\rightarrow\;
    \left\{\,\beta_j\ket{b_j}^{d_{i+1}}\,\right\}_{j=1}^k
\]
where $(d_i \leq D)$ and
$U_{i}(\alpha\ket{b})
=\textstyle\sum_j\beta_j\ket{b_j}$.

\textsc{Tagging} marks molecules that have reached the final depth and
are ready for aggregation:
\[
    \alpha\ket{b}^{d_{D}} \;\rightarrow\; \alpha~\dket{b}.
\]

\textsc{Interference} combines two tagged molecules that share a basis state,
performing coherent endpoint aggregation:
\[
    \alpha_1~\dket{b},\;\alpha_2~\dket{b}
    \;\rightarrow\;
    (\alpha_1+\alpha_2)~\dket{b}.
\]

This reaction is an amplitude-level operation, not a probability-level
calculation. Constructive interference increases the magnitude of the
endpoint aggregate, while destructive interference decreases it; in the
special case \(\alpha_1+\alpha_2=0\), the endpoint contribution is
annihilated and removed from the solution. Probabilities are assigned
only after aggregation has halted, by applying the Born rule to the
remaining endpoint aggregates.

A saturated execution applies these reactions until none remain,
recovering the same endpoint amplitudes as an ordinary state-vector
simulation. The key difference is that interference is now an explicit,
separately schedulable step rather than an implicit consequence of
matrix arithmetic. This separation is what makes threshold termination
possible, as described in the next section.

\section{Statistical Interference Sampling and Threshold Termination}
\label{sec:threshold}

Our approximation modifies only the saturated endpoint aggregation.
We introduce a threshold parameter~$\threshold$: once any tagged
endpoint molecule carries an amplitude whose magnitude reaches or
exceeds~$\threshold$,
the simulator halts, discards remaining reactions, forms one complex
aggregate per endpoint from the tracked sampling population, normalizes
the squared aggregate magnitudes, and samples from this partial
distribution. The method does not improve worst-case complexity. Its
purpose is a tunable tradeoff between the fraction of interference
reactions performed and the accuracy of the resulting output
distribution.

The procedure is summarized in Figure~\ref{alg:statistical-interference}.

\begin{figure}[htb]
\centering

\begin{minipage}{\dimexpr0.98\columnwidth-2\fboxsep\relax}
\footnotesize
\begin{algorithmic}[1]

\Require Circuit $U_1,\ldots,U_D$, input $|x_0\rangle$,
         threshold $\mathcal{T}$
\Ensure Sampled state $x$

\State $\mathcal{S} \gets \{\,1|x_0\rangle_{0}\,\}$
\State $\mathit{crossed} \gets \mathbf{false}$

\While{an enabled reaction remains}
    \State Select an enabled reaction from $\mathcal{S}$

    \If{reaction is time evolution}
        \State Select $\alpha|b\rangle_d \in \mathcal{S}$,
               $d < D$
        \State Compute
        $U_{d+1}(\alpha|b\rangle)
        = \sum_j \beta_j |b_j\rangle$
        \State Replace $\alpha|b\rangle_d$ by
        $\{\beta_j|b_j\rangle_{d+1}\}_j$

    \ElsIf{reaction is endpoint tagging}
        \State Replace $\alpha|b\rangle_D$ by tagged
               $\alpha|b\rangle$

    \ElsIf{reaction is interference}
        \State Select tagged
        $\alpha_1|b\rangle,\alpha_2|b\rangle \in \mathcal{S}$
        \State If the sum is zero, remove both molecules;
               otherwise replace them by
        $(\alpha_1+\alpha_2)|b\rangle$
    \EndIf

    \If{$\exists\,\alpha|b\rangle\in\mathcal{S}$ tagged with
         $|\alpha|\geq\mathcal{T}$}
        \State $\mathit{crossed} \gets \mathbf{true}$
        \State \textbf{break}
    \EndIf
\EndWhile

\State Discard unfinished reactions and untagged molecules
\State Let $A_b$ be the complex sum of amplitudes in the tracked
       sampling population at endpoint~$b$, including the trigger
\State $P(b) \gets
       |A_b|^2 /
       \sum_{b'} |A_{b'}|^2$
\State \Return $x \sim P$

\end{algorithmic}
\end{minipage}

\caption{Statistical interference sampling. Execution follows the
backend-induced order. The first tagged molecule
$\alpha\dket{b}$ satisfying $|\alpha|\geq\mathcal{T}$ that is admitted
by the serialized selection lock sets the stopping boundary. The
tracked population is then coherently aggregated by endpoint and
sampled from the normalized squared aggregate magnitudes.}
\label{alg:statistical-interference}
\end{figure}

The effectiveness of this tradeoff depends on the shape
of the output distribution. When a target endpoint’s amplitude is separated from all non-target endpoints by a gap exceeding the maximum attainable non-target amplitude, a threshold can be chosen that only a correct endpoint can cross. In that regime, the first molecule crossing the threshold must be the target, and the simulator returns the correct answer before full saturation. When the output distribution is flatter, non-target end-
points can cross the threshold before the final distribution resolves, and the approximation degrades accordingly.

A 4-qubit instance of Grover’s algorithm illustrates
both the promise and the difficulty. The standard textbook circuit contains 12 Hadamard gates and produces 4096 molecules in the solution, of which only 256 interfere constructively to form the correct output. The correct leaf states carry an average amplitude of $-0.003$; incorrect leaves carry an average amplitude of $\pm 0.0007$, as one can verify by tracing the amplitude arithmetic through the circuit. A threshold set between these values ensures that only the correct endpoint can trigger termination, since incorrect states are canceled by destructive interference before accumulating enough amplitude to cross it. At the same time, reaching that correct endpoint requires roughly 25,000 total interactions, of which only about 100 are the desired constructive ones. Even modest early termination, cutting 10\% of interactions, meaningfully reduces total computational effort, because the discarded reactions are overwhelmingly destructive. This analysis assumes a roughly uniform arrival order for molecules; the savings estimate depends on that assumption holding in practice. We formalize it in Appendix \ref{sec:formal} via a martingale-based bound on the probability that a zero-contribution endpoint triggers premature termination, and we measure how well it holds empirically across benchmark circuits in Section \ref{sec:empirical}.

\section{The Deutsch-Jozsa Example}
\label{sec:example}

We use the Deutsch-Jozsa \cite{Deutsch1985,deutschJozsa} circuit in Figure~\ref{fig:dj} as a simple example to illustrate our approach. This circuit determines whether a hidden function $f$ is constant or balanced by measuring the top wire as 0 or 1 respectively. A constant function, as the name suggests, returns identical output values for all possible input values, whereas a balanced function returns every value within a range with equal probability for any given input value. We first present its textbook simulation using vectors, then transition to our ChAM model.

\begin{figure}[htb]
  \begin{center}
    \begin{tikzpicture}
      \begin{yquant*}[operators/every barrier/.append style={red, thick}]
        qubit {} q[2];
        init {$\ket{0}$} q[0];
        init {$\ket{1}$} q[1];
        align -;
        h q[0];
        h q[1];
        align -;
        [y radius=0.46cm]
        box {$U_f$} (q[0,1]);
        h q[0];
        measure q[0];
      \end{yquant*}
    \end{tikzpicture}
  \end{center}
  \caption{\label{fig:dj} Instance of the Deutsch-Jozsa algorithm.}
\end{figure}

\subsection{State Vector Simulation}

In the standard Hilbert space model of quantum computing, a circuit with $n$ wires evolves a vector of size $2^n$, where each gate is represented as a $2^n \times 2^n$ unitary matrix~\cite{NielsenChuang2010}. The convention in representing vectors is that index $i$ (in binary) corresponds to the basis state $\ket{i}$. For example, the vector $(a~b~c~d)$ (transposed for convenience) represents the state $a \ket{00} + b \ket{01} + c \ket{10} + d \ket{11}$. 

Assuming the function inside the black-box unitary $U_f$ is the balanced function $f = \mathit{id}$, the gates used in the circuit have the following matrix representation:
\[
H = \frac{1}{\sqrt{2}}
\begin{pmatrix}
    1 & 1 \\ 1 & -1 
\end{pmatrix}
\qquad\qquad
U_f = 
\begin{pmatrix}
    1 & 0 & 0 & 0 \\
    0 & 1 & 0 & 0 \\
    0 & 0 & 0 & 1 \\
    0 & 0 & 1 & 0 
\end{pmatrix}
\]
Thus, starting with the initial state $\ket{01}$, the circuit evolves as follows:
\[
\begin{pmatrix} \sn{0} \\ \sn{1} \\ \sn{0} \\ \sn{0} \end{pmatrix}
\rightarrow
\begin{pmatrix} \sn{0} \\ \sn{0.7} \\ \sn{0} \\ \sn{0.7} \end{pmatrix}
\rightarrow
\begin{pmatrix} \sn{0.5} \\ \sn{-0.5} \\ \sn{0.5} \\ \sn{-0.5} \end{pmatrix}
\rightarrow
\begin{pmatrix} \sn{0.5} \\ \sn{-0.5} \\ \sn{-0.5} \\ \sn{0.5} \end{pmatrix}
\rightarrow
\begin{pmatrix} \sn{0} \\ \sn{0} \\ \sn{0.7} \\ \sn{-0.7} \end{pmatrix}
\]
Since the final state consists only of basis vectors $\ket{10}$ and~$\ket{11}$, measuring the first qubit always yields 1, correctly indicating that the hidden function is balanced and not constant. 

This example separates path generation from path recombination.  The first two Hadamard gates create the branches on which the oracle $U_f$ acts, but those branches are not by themselves the source of the algorithmic distinction.  The final Hadamard gate recombines paths with common measured endpoints: amplitudes for $\ket{00}$ and $\ket{01}$ cancel, while amplitudes for $\ket{10}$ and $\ket{11}$ reinforce.  In this sense, the final layer performs the endpoint-interference step that the ChAM model will later expose as a separate reaction.  

We can already observe a mathematical manifestation of these interference effects by expanding the intermediate steps in the final matrix-vector multiplication:
\[
\begin{pmatrix}
    \sn{0.7} & \sn{0} & \sn{0.7} & \sn{0} \\
    \sn{0} & \sn{0.7} & \sn{0} & \sn{0.7} \\
    \sn{0.7} & \sn{0} & \sn{-0.7} & \sn{0} \\
    \sn{0} & \sn{0.7} & \sn{0} & \sn{-0.7}
\end{pmatrix}
\begin{pmatrix} \sn{0.5} \\ \sn{-0.5} \\ \sn{-0.5} \\ \sn{0.5} \end{pmatrix}
= 
\begin{pmatrix} \sn{0.35-0.35} \\ \sn{-0.35+0.35} \\ \sn{0.35+0.35} \\ \sn{-0.35-0.35} \end{pmatrix}
= 
\begin{pmatrix} \sn{0} \\ \sn{0} \\ \sn{0.7} \\ \sn{-0.7} \end{pmatrix}
\]
where the next-to-last vector exposes the calculation of the interference effects: they correspond to summands coming from different indices of the matrix-vector multiplication. 

\begin{figure}[htb]
    \centering
    \includegraphics[width=0.8\linewidth]{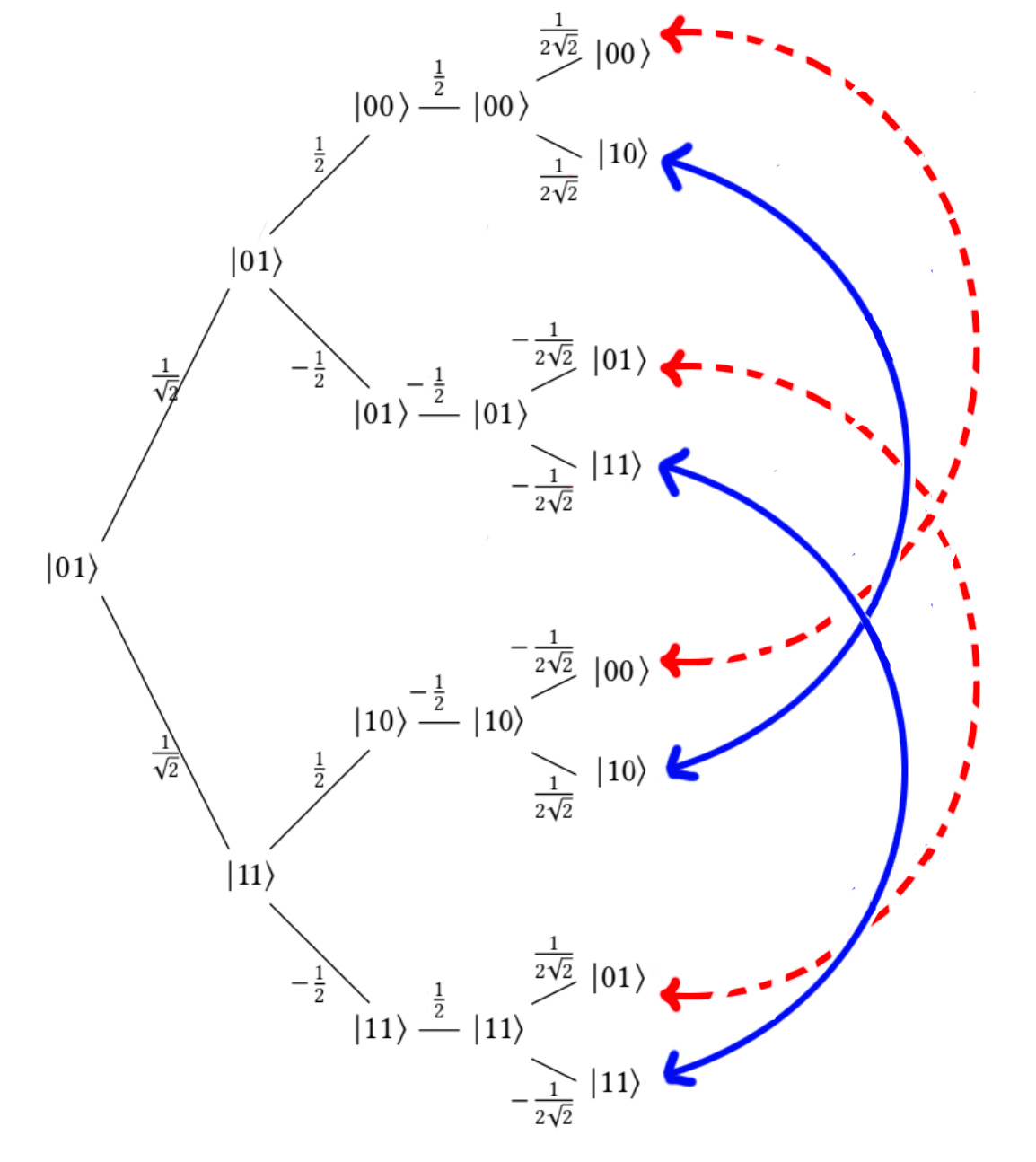}
    \caption{An abstract view of the ChAM model}
    \label{fig:abstractChAM}
\end{figure}

\subsection{ChAM Simulation}

\begin{figure*}[htb]
    \centering

    \subcaptionbox{}[0.3\textwidth]{%
        \centering
        \includegraphics[width=0.9\linewidth]{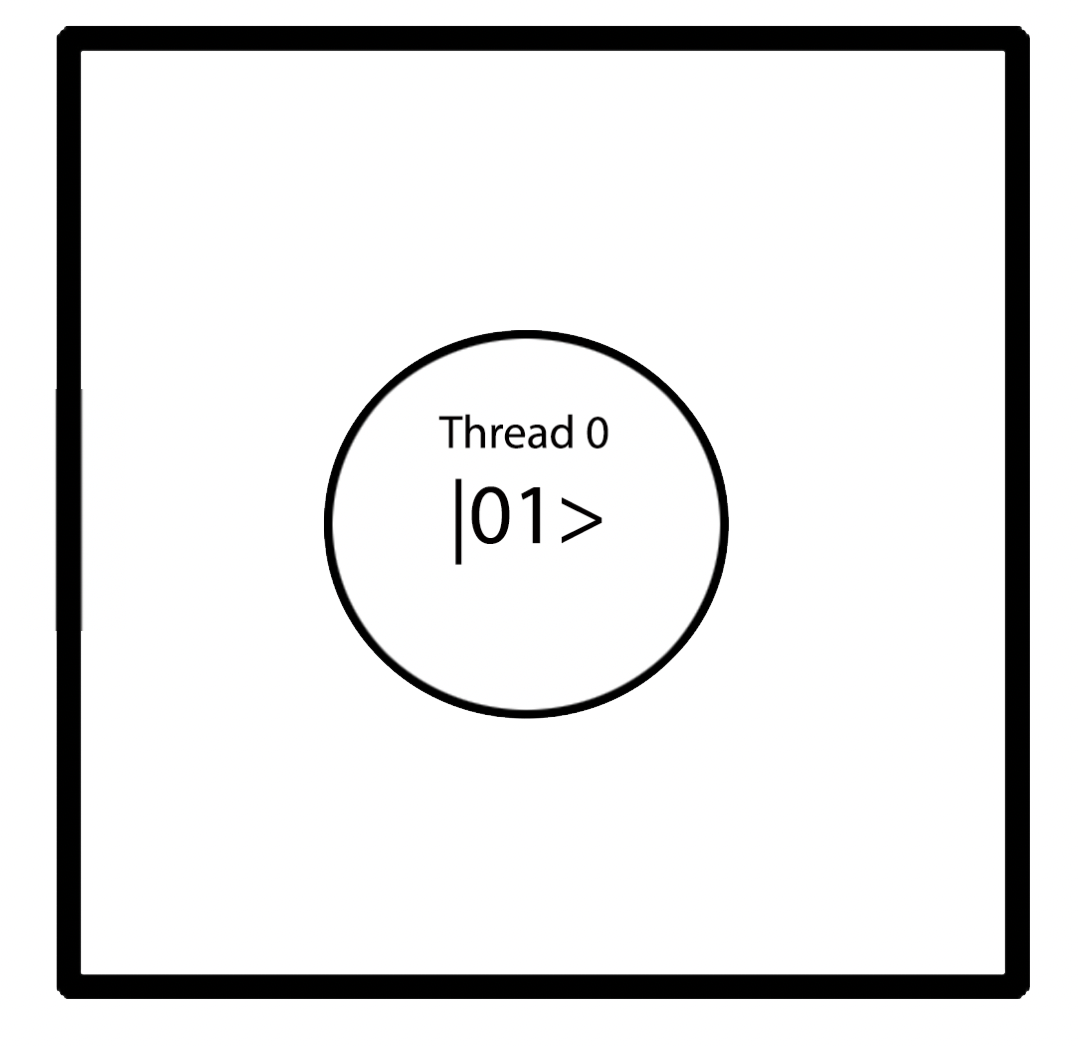}%
        \label{fig:cm1}%
    }\hfill
    \subcaptionbox{}[0.3\textwidth]{%
        \centering
        \includegraphics[width=0.9\linewidth]{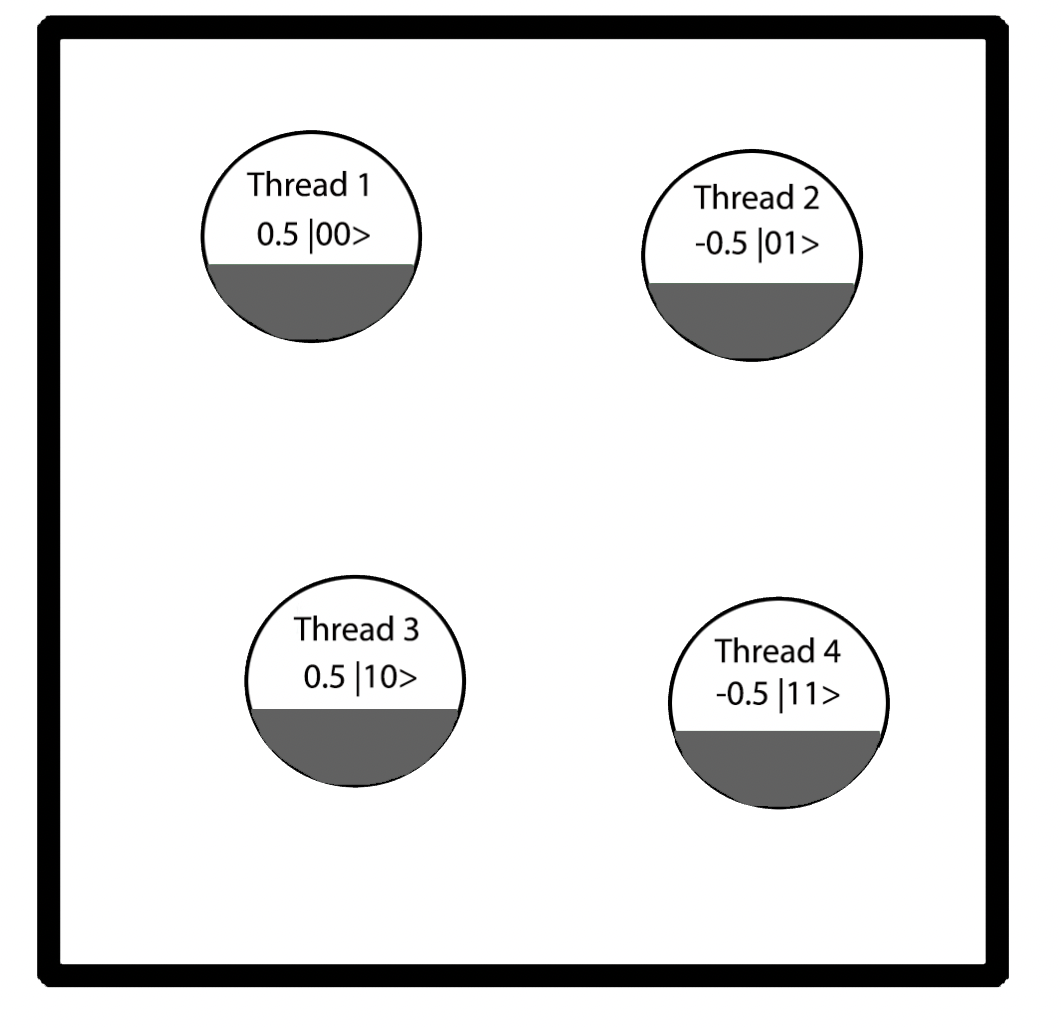}%
        \label{fig:cm2}%
    }\hfill
    \subcaptionbox{}[0.3\textwidth]{%
        \centering
        \includegraphics[width=0.9\linewidth]{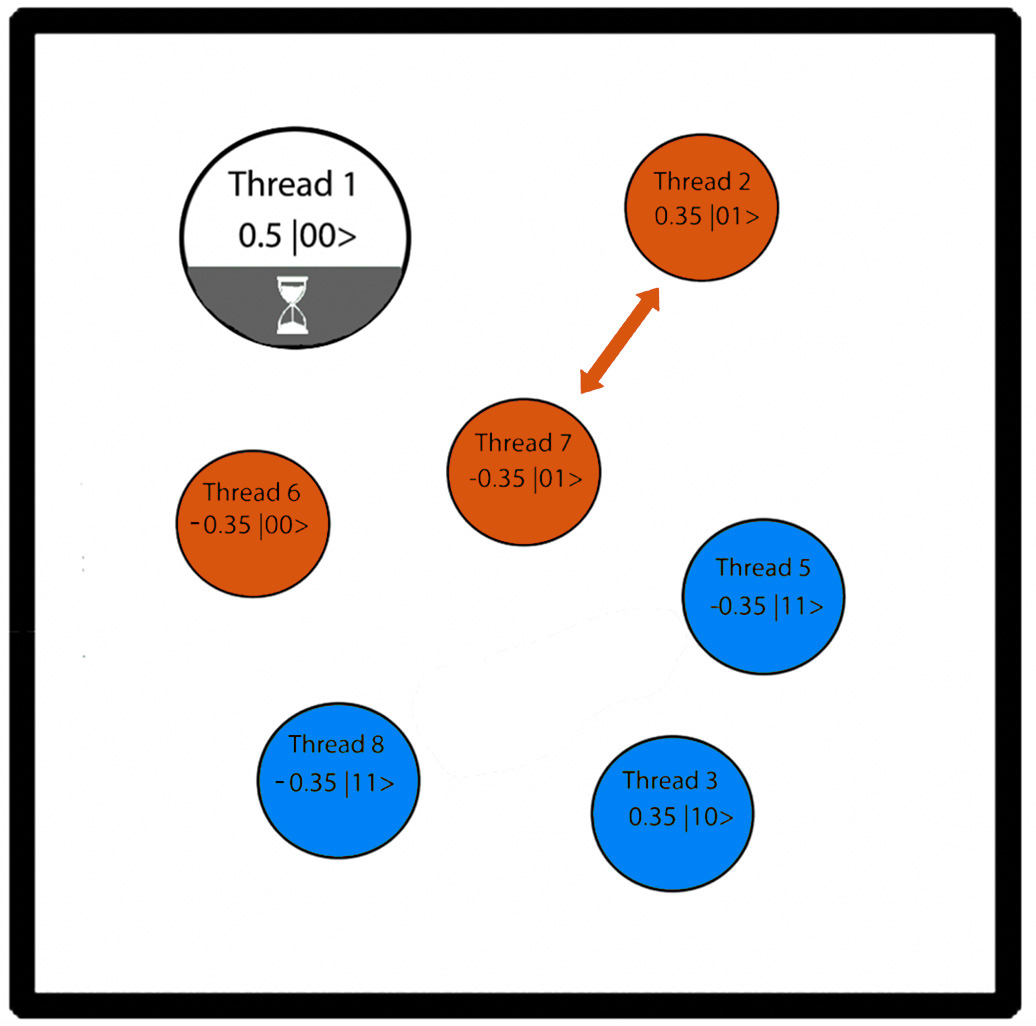}%
        \label{fig:cm3}%
    }

    \vspace{1em}

    \subcaptionbox{}[0.3\textwidth]{%
        \centering
        \includegraphics[width=0.9\linewidth]{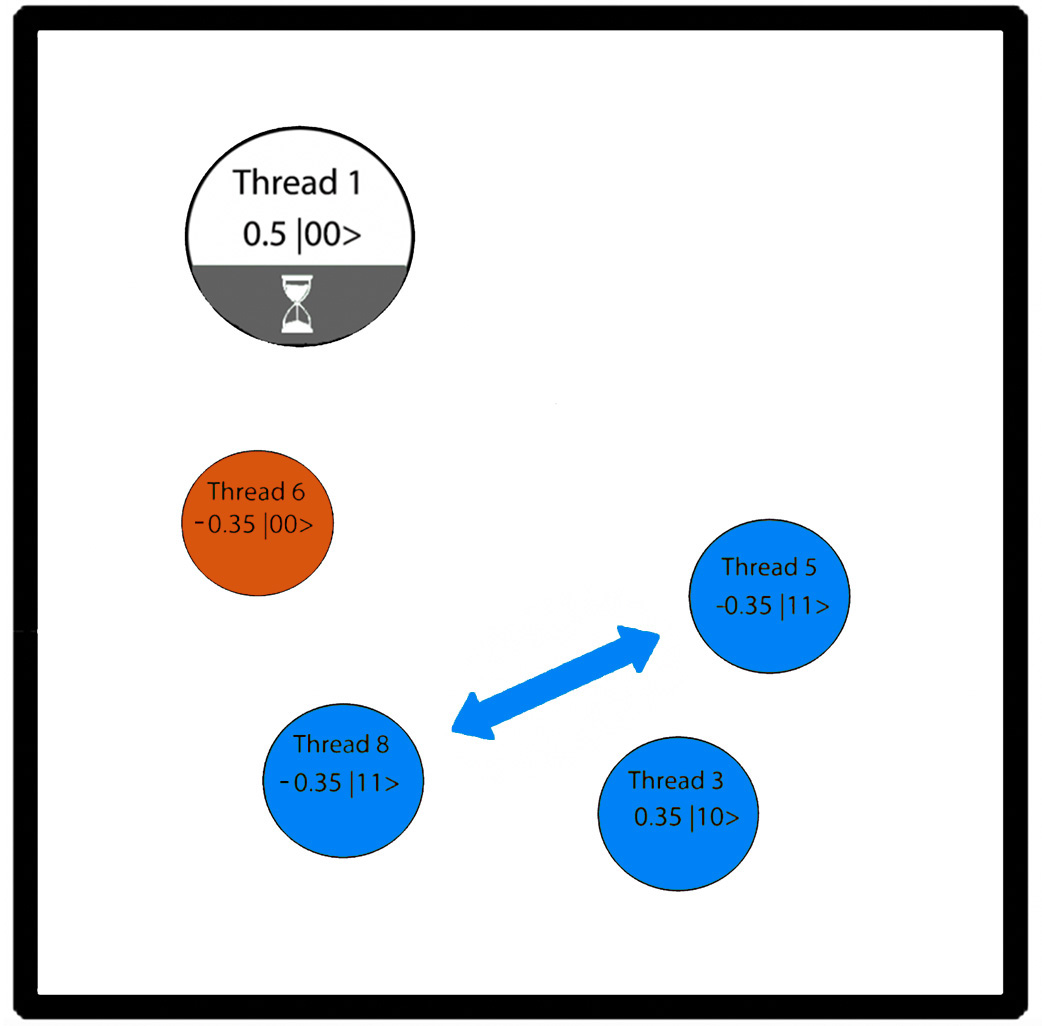}%
        \label{fig:cm4}%
    }\hfill
    \subcaptionbox{}[0.3\textwidth]{%
        \centering
        \includegraphics[width=0.9\linewidth]{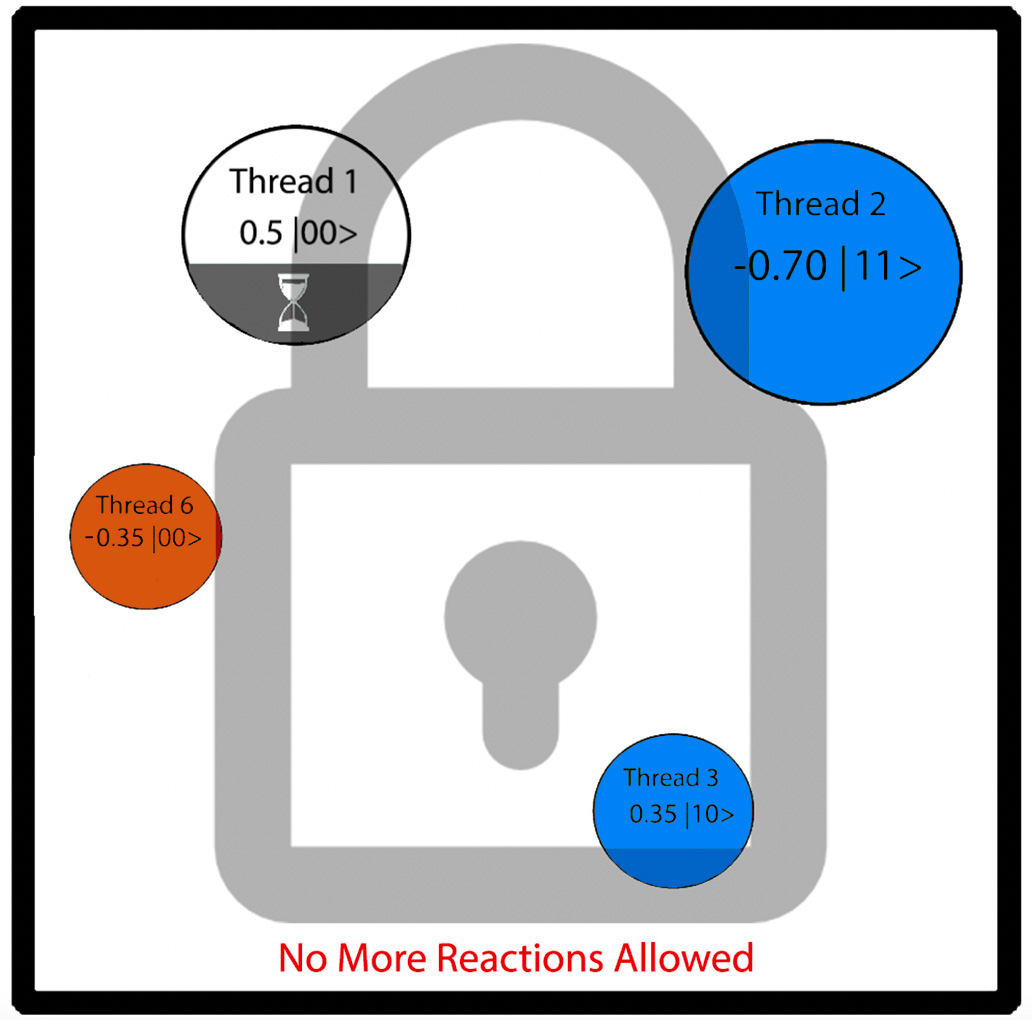}%
        \label{fig:cm5}%
    }

    \caption{Evolution of the chemical solution system throughout quantum interference simulation. 
    Each panel represents a key transformation step: 
    (a) Initial state, 
    (b) Hadamard gate application, 
    (c) Destructive interference of $\ket{01}$, 
    (d) Constructive interference preparation for $\ket{11}$, 
    and (e) Final state before sampling.}
    \label{fig:chemical_interference}
\end{figure*}

The same computation can be drawn as a path tree.  A gate that creates superposition branches the tree; with $h$ such branch points, the tree has up to $2^h$ leaves.  A direct state-vector simulation aggregates these leaves implicitly through matrix-vector multiplication.  In the ChAM simulation, by contrast, leaves are represented as asynchronously produced molecules, and endpoint interference is computed only when molecules with the same basis label react.  Figure~\ref{fig:abstractChAM} shows this logic for the Deutsch-Jozsa circuit.  The computation begins at $\ket{01}$; each Hadamard gate branches the execution; the oracle acts locally on each branch; and the final Hadamard layer produces terminal molecules that must be combined.  Blue arrows indicate reinforcing endpoint contributions, and red arrows indicate canceling contributions.

The actual ChAM execution relaxes the strict tree ordering shown in the figure.  Molecules may be produced and combined under different concurrent schedules, and endpoint aggregation can begin before every branch has finished.  This is the point at which thresholding enters: if a tagged molecule's amplitude magnitude reaches or exceeds the chosen threshold, the simulator can stop before all remaining contributions have been processed.

Figure~\ref{fig:chemical_interference} gives a more concrete execution trace.  Initially, a single molecule carries $\ket{01}$ with amplitude $1$.  After the first Hadamard layer, the solution contains several partially evolved molecules; once a molecule reaches the final depth, it is marked as eligible for interference by coloring it. For clarity, we use two colors as follows: blue denotes endpoints that would survive in the full saturated Deutsch-Jozsa calculation; orange denotes endpoints that would be canceled by a full saturated calculation.  To illustrate pruning, the interference calculation for one orange molecule carrying $\ket{00}$ is delayed since its counterpart with the grey shading has not reached the final depth.  The other completed molecules can still react: two $\ket{01}$ molecules cancel destructively, while two contributions to~$\ket{11}$ combine constructively. As the resulting amplitude for $\ket{11}$ exceeds the threshold of 0.5, the simulation stops and samples from the partial marked solution that includes $\ket{00}$, $\ket{10}$, and $\ket{11}$.  The example therefore shows both the benefit and the approximation error of the method: useful constructive interference can be observed early, but the unprocessed molecules can leave residual probability on an endpoint that would disappear in the saturated calculation.

\section{Benchmark Algorithms and Experimental Setup}
\label{sec:background}

We evaluate the method on three standard families of algorithms
chosen to span a range of output-distribution profiles, from strongly
peaked to comparatively flat. For each algorithm we give a brief
formal description and highlight the interference characteristics
most relevant to our evaluation.

Let $[2^n]=\{0,1,\ldots,2^n-1\}$.
In Simon's problem~\cite{doi:10.1137/S0097539796298637}, one is
given a two-to-one function $f:[2^n]\rightarrow[2^n]$ with the
promise that there is a secret string $a\in[2^n]$ such that
$f(x)=f(x\oplus a)$ for all $x$; the goal is to infer $a$. The
algorithm works by preparing a superposition, applying~$f$, and
measuring, producing a random element of the dual space of $a$;
repeated runs recover $a$ by solving a linear system. In Grover
search~\cite{10.1145/237814.237866}, one is given an oracle
$f:[2^n]\rightarrow[2]$ with a unique marked input $u$ satisfying
$f(u)=1$; the goal is to sample $u$ with high probability. The
algorithm amplifies the amplitude of $u$ through repeated
inversion-about-the-mean operations, concentrating probability on
a single endpoint after $O(\sqrt{2^n})$ iterations. In Shor period
finding~\cite{doi:10.1137/S0036144598347011}, one considers a
periodic modular function $f(x)=a^x \bmod N$ and seeks the period
of $f$. The algorithm applies a quantum Fourier transform to extract
period information encoded in the amplitude distribution, producing
output spread across multiple valid frequencies rather than a single
marked state.

These three families together span the interference regimes most
relevant to our evaluation. All three circuit families share the same
path-sum structure. Hadamard layers generate a large number of
branches, oracle or modular-exponentiation blocks attach
problem-dependent phases or values, and final interference layers
concentrate probability on endpoints with the desired algebraic
structure. In the ChAM model, the branching layers create many
molecules, the oracle updates their local basis states and amplitudes,
and the final layers determine which endpoint molecules are likely to
combine constructively before the threshold is reached.

\subsection{Scheduling Effects and the Gap Between Model and Practice}
\label{sec:scheduling}

Establishing error bounds for statistical approximation is valuable. In Appendix~\ref{sec:formal} we derive a bound on the probability that a
zero-contribution endpoint triggers early termination under a uniform distribution model. The implementation, however, realizes a different distribution over arrival orders. Each path contribution is submitted as an independent unit of work, and the language runtime and operating system multiplex many runnable tasks over a finite number of hardware execution contexts. This is neither uniform nor
fair in the relevant sense. The resulting completion order depends on implementation-level effects such as work-queue policy, preemption, cache locality, memory contention, and, on heterogeneous processors, assignment to cores with different performance characteristics. None of these mechanisms is designed to sample path contributions uniformly. The observed schedule should therefore be understood as an implementation-dependent distribution over arrival orders, whereas the formal analysis studies an idealized random-permutation model. The run-to-run variation at fixed threshold values partly reflects this distinction.

Several factors contribute to this unfairness. First, our
implementation maps each path contribution to a concurrent thread, and
in our experimented machines the OS kernel uses an $m$:$n$ threading model in which $m$ application
threads are multiplexed onto $n$ physical processor cores. When $m > n$
the kernel must decide which threads run when, and it does so based on
cost and priority heuristics rather than any notion of fairness across
path contributions. Since all branches of our simulation compute the
same sequence of gates on different basis states, they all have
approximately the same computational cost, giving the scheduler no
signal to prioritize one over another. The resulting order is
effectively determined by low-level timing effects. Second, modern
processors typically include two classes of cores: high-performance
cores (P-cores) optimized for speed, and efficiency cores (E-cores)
optimized for power consumption. A thread scheduled on a P-core will
complete its work faster than one on an E-core, meaning that which
path contributions arrive first depends in part on which physical core
each thread happens to be assigned to, which is an assignment the application
cannot control. Together, these effects mean that the arrival order
observed in practice can differ substantially from the uniform random
permutation assumed by the formal model. Establishing a rigorous
probabilistic model of the execution order under realistic scheduling
conditions is outside the scope of this work. The run-to-run
variability in accuracy observed at fixed threshold values in the
results above reflects this gap in part: it is a consequence of
scheduling nondeterminism as much as of the approximation method
itself, and any future theoretical analysis will need to account
for it.

\subsection{Experimental Protocol and Metrics}
\label{sec:experimental-protocol}

For each threshold and output-selection configuration, we perform
50 independent fresh-JVM trials. We organize the reported quantities
starting with the basic terminal-contribution and interference-work
counts, followed by task-level success and distributional fidelity.

We begin with the basic terminal-contribution counts.
$B_{\mathrm{total}}$ is the total number of terminal path
contributions generated by the corresponding configuration,
$B_{\mathrm{correct}}$ is the number associated with accepted
outputs, and $B_{\mathrm{active}}$ is the number of terminal
contributions that have reached the terminal selection boundary when
sampling occurs. We additionally record
$B_{\mathrm{active,correct}}$, the corresponding number associated
with accepted outputs. These quantities characterize how far path
generation and endpoint arrival have progressed, but they are not
themselves counts of completed interference operations.

We therefore measure endpoint-interference work directly.
Let $B^{(I)}_x$ denote the number of terminal contributions admitted
by the terminal policy at endpoint $x$ in the saturated execution. We
use the following canonical saturated binary-reduction reference
count:
\[
  \text{$\displaystyle
  I_{\mathrm{full}}
  = \sum_x \max\!\left(B^{(I)}_x-1,0\right)
  $}.
\]
Let $I_{\mathcal{T}}$ be the number of compatible endpoint-addition
reactions completed when execution halts. We define the interference reduction
relative to this canonical reference as
\[
  \text{$\displaystyle
  R_{\mathcal{T}}
  = 1-\frac{I_{\mathcal{T}}}{I_{\mathrm{full}}}
  $}.
\]
This measure is generally not equal to
$1-B_{\mathrm{active}}/B_{\mathrm{total}}$. It measures one internal
operation class relative to the stated reference. We report
$B_{\mathrm{active}}$ and $I_{\mathcal{T}}$ as means $\pm$ sample
standard deviations and give a $95\%$ confidence interval for
$R_{\mathcal{T}}$.

As a statistical check on work reduction, we compare
$I_{\mathcal{T}}$ with $I_{\mathrm{full}}$ using both a paired
$t$-test and a Wilcoxon signed-rank test. For every reported
configuration included in this comparison, both tests reject the
null hypothesis of no reduction. The Wilcoxon tests give
$p<8\times10^{-10}$ in every configuration, and the largest two-sided
paired-$t$-test $p$-value is below $6.52\times10^{-39}$. Because both
tests are highly sensitive at the observed effect sizes, we emphasize
the reduction fraction and its confidence interval rather than
significance alone

We evaluate two output-selection policies. Under normalized-Born sampling, the simulation stops when the
first end-point reaches threshold $\threshold$, but the output is sampled from the normalized Born distribution of the end-
point aggregates available at that time. Under early termination, labeled ``Early'', the endpoint that first reaches $\threshold$ is returned
directly.

\textit{Task-level success} is the fraction of trials that
return a correct or valid result under the algorithm-specific
acceptance criterion. We report the observed proportion with a
$95\%$ Wilson confidence interval. For Grover search, success means
selecting the marked state or satisfying assignment. For Simon's
algorithm with hidden string $a$, a valid sampled string $y$ satisfies
$y\mathbin{\cdot}a=0\pmod 2$. For Shor's algorithm, success means
that classical post-processing constructs non-trivial factors of $N$.

To quantify distributional fidelity, we additionally measure
the total-variation distance
\[
  \text{$\displaystyle
  \delta_{\mathrm{TV}}(Q_{\mathcal{T}},P^*)
  = \frac{1}{2}\sum_x
  \left|Q_{\mathcal{T}}(x)-P^*(x)\right|
  $},
\]
where $Q_{\mathcal{T}}$ is the empirical distribution of
outputs from the thresholded trials and $P^*$ is the exact saturated
distribution. The reported TVD therefore includes both approximation
error and finite-sample variability. We estimate its uncertainty
using $5{,}000$ bootstrap replicates. TVD is retained as the primary measure as it
is also an error metric used in Appendix \ref{sec:formal}

As a complementary distributional metric, we also compute the
Hellinger distance
\[
  \text{$\displaystyle
  H(Q_{\mathcal T},P^*)
  = \frac{1}{\sqrt{2}}
  \left[\sum_x\left(
  \sqrt{Q_{\mathcal T}(x)}-\sqrt{P^*(x)}
  \right)^2\right]^{1/2}
  $},
\]
together with
\[
  \text{$\displaystyle
  F_H(Q_{\mathcal T},P^*)=1-H(Q_{\mathcal T},P^*)^2
  $},
\]
which we call the Hellinger affinity. Hellinger results are
reported in Appendix \ref{app:hellinger} as a complementary check
on the qualitative trends identified by TVD.

For the early termination policy we additionally record the
amplitude separation present when the selection is made. Let
$A_{\mathrm{sel}}$ denote the aggregate amplitude of the
selected endpoint and let
\[
 \text{$\displaystyle 
 A_{\mathrm{inc}}^{\max}
  =
  \max_{i\notin C}
  |A_i^{\mathrm{ready}}| $}
\]
denote the largest aggregate amplitude among incorrect
endpoints represented in the instrumented ready pool at the
sampling boundary. We define the empirical
\textit{selection-separation margin}
\[ \text{$\displaystyle 
  M_{\mathrm{sel}}
  =
  |A_{\mathrm{sel}}|
  -
  A_{\mathrm{inc}}^{\max} $}
\]
This quantity is an execution-time analogue of the
threshold-resolvable amplitude gap introduced in
Appendix~A. A positive value means that the selected endpoint
is separated in magnitude from every incorrect endpoint
visible in the ready-pool snapshot at that time.

Most experiments were run on an Apple MacBook Pro with 64GB
RAM and an M2 Max processor; larger Grover and Shor instances
were run on an AWS m8g.48xlarge instance (192 vCPUs, 768 GB
RAM) because of memory constraints.

\section{Empirical Evaluation}
\label{sec:empirical}

\subsection{Simulating Grover's Algorithm}
\label{sec:grover}

We consider Grover search instances ranging from 4 to 8 qubits. The
marked item is either an explicitly tagged basis state, such as
$\ket{000}$, $\ket{110}$, or $\ket{010}$, or the satisfying
assignment of a small 3-SAT instance. We also test the implementation on a 30-qubit tagged-state
instance to extend the evaluation to a larger configuration.Figure~\ref{fig:grovercorr}
reports the fraction of runs in which the thresholded simulation
returns the marked output across a range of threshold values.
The saturated reference is the unthresholded execution in which
all compatible endpoint reactions are allowed to complete.

\begin{table*}[htb]
\centering
\caption{
Grover results under normalized-Born (Born) and
early-selection (Early) output.
$B_{\mathrm{active}}$ is the number of terminal contributions active
when output selection occurs. $B_c/B=B_{\mathrm{correct}}/
B_{\mathrm{total}}$, where $B_{\mathrm{correct}}$ is the saturated
terminal population associated with the correct output and
$B_{\mathrm{total}}$ is the total saturated terminal population.
$I_{\mathcal{T}}$ is the number of compatible endpoint-interference
reactions completed before selection, and $I_{\mathrm{full}}$ is the
corresponding saturated count. Reduction is
$1-I_{\mathcal{T}}/I_{\mathrm{full}}$. $B_{\mathrm{active}}$ and
$I_{\mathcal{T}}$ are reported as mean $\pm$ sample standard
deviation. Reduction and task success are reported with $95\%$
confidence intervals. TVD is the empirical total-variation distance
from the exact saturated distribution.
}
\label{tab:grover-results}

\scriptsize
\setlength{\tabcolsep}{2.5pt}
\renewcommand{\arraystretch}{1.08}

\resizebox{\dimexpr\textwidth-2\fboxsep\relax}{!}{%
\begin{tabular}{lllrrrrrr}
\toprule
Instance &
Policy &
$\mathcal T$ &
$B_{\mathrm{active}}$ &
$B_c/B$ &
$I_{\mathcal T}/I_{\mathrm{full}}$ &
Reduction (\%) &
Success (\%) [95\% CI] &
TVD \\
\midrule


\multirow{10}{*}{Tagged Grover (4q)}
& \multirow{5}{*}{Born}
& 0.1
& $103.6\pm13.4$
& $256/4096$
& $72.1\pm11.7\,/\,4080$
& $98.23\,[98.15,98.32]$
& $28\,[17.47,41.67]$
& 0.614 \\

&
& 0.3
& $738.8\pm98.6$
& $256/4096$
& $1169.0\pm97.6\,/\,4080$
& $71.35\,[70.67,72.03]$
& $44\,[31.16,57.69]$
& 0.384 \\

&
& 0.5
& $1209.8\pm50.3$
& $256/4096$
& $2078.2\pm49.3\,/\,4080$
& $49.06\,[48.72,49.41]$
& $67\,[53.17,78.41]$
& 0.379 \\

&
& 0.7
& $1433.5\pm121.9$
& $256/4096$
& $1546.9\pm122.0\,/\,4080$
& $62.09\,[61.65,62.50]$
& $78\,[64.76,87.25]$
& 0.409 \\

&
& 0.9
& $3176.8\pm14.6$
& $256/4096$
& $2024.0\pm20.9\,/\,4080$
& $50.39\,[50.31,50.45]$
& $94\,[83.78,97.94]$
& 0.324 \\

\cmidrule(lr){2-9}

& \multirow{5}{*}{Early}
& 0.1
& $82.6\pm16.5$
& $256/4096$
& $71.1\pm13.6\,/\,4080$
& $98.26\,[98.16,98.35]$
& $34\,[22.44,47.85]$
& 0.930 \\

&
& 0.3
& $541.8\pm74.0$
& $256/4096$
& $1200.9\pm74.5\,/\,4080$
& $70.57\,[70.05,71.08]$
& $96\,[86.54,98.90]$
& 0.492 \\

&
& 0.5
& $918.9\pm51.9$
& $256/4096$
& $2077.4\pm50.5\,/\,4080$
& $49.08\,[48.73,49.43]$
& $100\,[92.87,100.0]$
& 0.527 \\

&
& 0.7
& $1166.9\pm117.6$
& $256/4096$
& $1544.3\pm116.5\,/\,4080$
& $62.15\,[61.74,62.56]$
& $100\,[92.87,100.0]$
& 0.527 \\

&
& 0.9
& $2120.7\pm14.7$
& $256/4096$
& $2018.1\pm20.8\,/\,4080$
& $50.54\,[50.45,50.60]$
& $100\,[92.87,100.0]$
& 0.527 \\

\midrule


\multirow{10}{*}{3-SAT Grover (8q)}
& \multirow{5}{*}{Born}
& 0.1
& $1307.7\pm25.0$
& $32768/262144$
& $500.1\pm22.6\,/\,32736$
& $98.47\,[98.45,98.49]$
& $34\,[22.44,47.85]$
& 0.696 \\

&
& 0.3
& $13107.3\pm19.8$
& $32768/262144$
& $12736.1\pm21.8\,/\,32736$
& $61.09\,[61.08,61.11]$
& $51\,[37.58,64.28]$
& 0.322 \\

&
& 0.5
& $132344.7\pm90.7$
& $32768/262144$
& $32191.1\pm88.8\,/\,32736$
& $1.66\,[1.59,1.74]$
& $63\,[49.14,75.00]$
& 0.314 \\

&
& 0.7
& $157310.4\pm141.2$
& $32768/262144$
& $21592.5\pm139.3\,/\,32736$
& $34.04\,[33.97,34.10]$
& $86\,[73.81,93.05]$
& 0.442 \\

&
& 0.9
& $169775.2\pm24.2$
& $32768/262144$
& $30676.7\pm25.9\,/\,32736$
& $6.29\,[5.83,5.85]$
& $96\,[86.54,98.90]$
& 0.309 \\

\cmidrule(lr){2-9}

& \multirow{5}{*}{Early}
& 0.1
& $628.2\pm24.1$
& $32768/262144$
& $499.6\pm22.4\,/\,32736$
& $98.47\,[98.45,98.49]$
& $34\,[22.44,47.85]$
& 0.976 \\

&
& 0.3
& $12885.7\pm20.3$
& $32768/262144$
& $12734.2\pm21.1\,/\,32736$
& $61.10\,[61.08,61.12]$
& $100\,[92.87,100.0]$
& 0.742 \\

&
& 0.5
& $32341.9\pm75.2$
& $32768/262144$
& $32186.0\pm76.5\,/\,32736$
& $1.68\,[1.61,1.75]$
& $100\,[92.87,100.0]$
& 0.742 \\

&
& 0.7
& $43290.5\pm201.3$
& $32768/262144$
& $21589.7\pm201.1\,/\,32736$
& $34.05\,[33.97,34.14]$
& $100\,[92.87,100.0]$
& 0.742 \\

&
& 0.9
& $61774.7\pm22.9$
& $32768/262144$
& $30824.8\pm21.9\,/\,32736$
& $5.84\,[5.83,5.85]$
& $100\,[92.87,100.0]$
& 0.742 \\

\midrule


\multirow{10}{*}{Grover (30q)}
& \multirow{5}{*}{Born}
& 0.1
& $251078.4\pm4800.0$
& $196608/50331648$
& $96019.2\pm4339.2\,/\,6285312$
& $98.47\,[98.45,98.49]$
& $30\,[24.44,38.85]$
& 0.896 \\

&
& 0.3
& $2516601.6\pm3801.6$
& $196608/50331648$
& $3475331.2\pm3186.6\,/\,6285312$
& $44.71\,[61.08,61.11]$
& $49\,[40.58,60.28]$
& 0.687 \\

&
& 0.5
& $25410182.4\pm17414.4$
& $196608/50331648$
& $3870691.2\pm7049.6\,/\,6285312$
& $38.42\,[1.59,1.74]$
& $63\,[54.14,72.00]$
& 0.413 \\

&
& 0.7
& $30203596.8\pm27110.4$
& $196608/50331648$
& $4574460.0\pm6745.6\,/\,6285312$
& $27.22\,[33.97,34.10]$
& $85\,[77.81,93.05]$
& 0.548 \\

&
& 0.9
& $32596838.4\pm4646.4$
& $196608/50331648$
& $5889926.4\pm14972.8\,/\,6285312$
& $6.29\,[6.27,6.31]$
& $91\,[87.54,96.90]$
& 0.401 \\

\cmidrule(lr){2-9}

& \multirow{5}{*}{Early}
& 0.1
& $120614.4\pm4627.2$
& $196608/50331648$
& $85923.2\pm1300.8\,/\,6285312$
& $98.63\,[98.45,98.49]$
& $34\,[20.44,47.85]$
& 0.996 \\

&
& 0.3
& $2474054.4\pm3897.6$
& $196608/50331648$
& $444966.4\pm2051.2\,/\,6285312$
& $92.92\,[61.08,61.12]$
& $100\,[94.87,100.0]$
& 0.942 \\

&
& 0.5
& $6209644.8\pm14438.4$
& $196608/50331648$
& $2179712.0\pm6488.0\,/\,6285312$
& $65.32\,[1.61,1.75]$
& $100\,[94.87,100.0]$
& 0.773 \\

&
& 0.7
& $8311776.0\pm38649.6$
& $196608/50331648$
& $4445222.4\pm8361.2\,/\,6285312$
& $29.28\,[33.97,34.14]$
& $100\,[94.87,100.0]$
& 0.773 \\

&
& 0.9
& $11860742.4\pm4396.8$
& $196608/50331648$
& $4918361.6\pm4204.8\,/\,6285312$
& $21.75\,[5.83,6.85]$
& $100\,[94.87,100.0]$
& 0.773 \\

\bottomrule
\end{tabular}
}

\end{table*}

\begin{figure*}[htb]
\centering

\begin{tikzpicture}


\begin{axis}[
    name=groverA,
    width=0.45\textwidth,
    height=0.33\textwidth,
    title={(a) Correct-output rate},
    xlabel={\threshold},
    ylabel={Correct outputs (\%)},
    xmin=0.08,
    xmax=0.92,
    ymin=0,
    ymax=105,
    xtick={0.1,0.2,0.3,0.4,0.5,0.6,0.7,0.8,0.9},
    ytick={0,20,40,60,80,100},
    grid=major,
    tick label style={font=\scriptsize},
    label style={font=\small},
    title style={font=\small},
    legend to name=groverlegend,
    legend columns=2,
    legend style={
        font=\scriptsize,
        draw=none,
        /tikz/every even column/.append style={column sep=0.8em}
    }
]

\addplot+[
    color=cbBlue,
    mark=*,
    solid,
    thick,
    error bars/.cd,
        y dir=both,
        y explicit
] table[
    x=T,
    y=mean,
    y error minus=emin,
    y error plus=eplus
] {
T   mean   emin   eplus
0.1 28     10.53  13.67
0.2 38     12.14  13.85
0.3 44     12.84  13.69
0.4 59     13.80  12.51
0.5 67     13.83  11.41
0.6 74     13.55  10.13
0.7 78     13.24   9.25
0.8 89     11.59   6.03
0.9 94     10.22   3.94
};
\addlegendentry{Tagged state, Born}

\addplot+[
    color=cbOrange,
    mark=square*,
    solid,
    thick,
    error bars/.cd,
        y dir=both,
        y explicit
] table[
    x=T,
    y=mean,
    y error minus=emin,
    y error plus=eplus
] {
T   mean   emin   eplus
0.1 34     11.56  13.85
0.2 49     13.28  13.42
0.3 51     13.42  13.28
0.4 57     13.73  12.73
0.5 63     13.86  12.00
0.6 77     13.33   9.48
0.7 86     12.19   7.05
0.8 91     11.11   5.26
0.9 96      9.46   2.90
};
\addlegendentry{3-SAT, Born}

\addplot+[
    color=green,
    mark=square*,
    solid,
    thick,
    error bars/.cd,
        y dir=both,
        y explicit
] table[
    x=T,
    y=mean,
    y error minus=emin,
    y error plus=eplus
] {
T   mean   emin   eplus
0.1 30     5.56  8.85
0.2 42     8.28  11.42
0.3 49     8.42  11.28
0.4 52     8.73  9.73
0.5 63     8.86  9.00
0.6 73     8.33   8.48
0.7 85     7.19   8.05
0.8 87     5.11   6.26
0.9 91      3.46   5.90
};
\addlegendentry{30-qubit tagged, Born}

\addplot+[
    color=cbBlue,
    mark=triangle*,
    dashed,
    thick,
    error bars/.cd,
        y dir=both,
        y explicit
] table[
    x=T,
    y=mean,
    y error minus=emin,
    y error plus=eplus
] {
T   mean   emin   eplus
0.1 34     11.56  13.85
0.2 95      9.86   3.44
0.3 96      9.46   2.90
0.4 100     7.13   0
0.5 100     7.13   0
0.6 100     7.13   0
0.7 100     7.13   0
0.8 100     7.13   0
0.9 100     7.13   0
};
\addlegendentry{Tagged state, early selection}

\addplot+[
    color=cbOrange,
    mark=diamond*,
    dashed,
    thick,
    error bars/.cd,
        y dir=both,
        y explicit
] table[
    x=T,
    y=mean,
    y error minus=emin,
    y error plus=eplus
] {
T   mean   emin   eplus
0.1 34     11.56  13.85
0.2 100     7.13   0
0.3 100     7.13   0
0.4 100     7.13   0
0.5 100     7.13   0
0.6 100     7.13   0
0.7 100     7.13   0
0.8 100     7.13   0
0.9 100     7.13   0
};
\addlegendentry{3-SAT, early selection}

\addplot+[
    color=green,
    mark=diamond*,
    dashed,
    thick,
    error bars/.cd,
        y dir=both,
        y explicit
] table[
    x=T,
    y=mean,
    y error minus=emin,
    y error plus=eplus
] {
T   mean   emin   eplus
0.1 34     13.56  13.85
0.2 36     5.13   0
0.3 100     5.13   0
0.4 100     5.13   0
0.5 100     5.13   0
0.6 100     5.13   0
0.7 100     5.13   0
0.8 100     5.13   0
0.9 100     5.13   0
};
\addlegendentry{30-qubit tagged, early selection}

\end{axis}


\begin{axis}[
    name=groverB,
    at={(groverA.east)},
    anchor=west,
    xshift=1.45cm,
    width=0.45\textwidth,
    height=0.33\textwidth,
    title={(b) Distance from saturated output},
    xlabel={\threshold},
    ylabel={TVD from saturated distribution},
    xmin=0.08,
    xmax=0.92,
    ymin=0,
    ymax=1,
    xtick={0.1,0.2,0.3,0.4,0.5,0.6,0.7,0.8,0.9},
    ytick={0,0.2,0.4,0.6,0.8,1.0},
    grid=major,
    tick label style={font=\scriptsize},
    label style={font=\small},
    title style={font=\small}
]

\addplot+[
    color=cbBlue,
    mark=*,
    solid,
    thick
] coordinates {
    (0.1,0.614)
    (0.2,0.749)
    (0.3,0.384)
    (0.4,0.334)
    (0.5,0.379)
    (0.6,0.449)
    (0.7,0.409)
    (0.8,0.339)
    (0.9,0.324)
};

\addplot+[
    color=cbOrange,
    mark=square*,
    solid,
    thick
] coordinates {
    (0.1,0.696)
    (0.2,0.333)
    (0.3,0.322)
    (0.4,0.366)
    (0.5,0.314)
    (0.6,0.362)
    (0.7,0.442)
    (0.8,0.414)
    (0.9,0.309)
};

\addplot+[
    color=green,
    mark=square*,
    solid,
    thick
] coordinates {
    (0.1,0.896)
    (0.2,0.733)
    (0.3,0.687)
    (0.4,0.666)
    (0.5,0.413)
    (0.6,0.236)
    (0.7,0.548)
    (0.8,0.519)
    (0.9,0.401)
};

\addplot+[
    color=cbBlue,
    mark=triangle*,
    dashed,
    thick
] coordinates {
    (0.1,0.930)
    (0.2,0.965)
    (0.3,0.492)
    (0.4,0.527)
    (0.5,0.527)
    (0.6,0.527)
    (0.7,0.527)
    (0.8,0.527)
    (0.9,0.527)
};

\addplot+[
    color=cbOrange,
    mark=diamond*,
    dashed,
    thick
] coordinates {
    (0.1,0.976)
    (0.2,0.742)
    (0.3,0.742)
    (0.4,0.742)
    (0.5,0.742)
    (0.6,0.742)
    (0.7,0.742)
    (0.8,0.742)
    (0.9,0.742)
};

\addplot+[
    color=green,
    mark=diamond*,
    dashed,
    thick
] coordinates {
    (0.1,0.996)
    (0.2,0.972)
    (0.3,0.942)
    (0.4,0.773)
    (0.5,0.773)
    (0.6,0.773)
    (0.7,0.773)
    (0.8,0.773)
    (0.9,0.773)
};

\end{axis}

\begin{pgfonlayer}{background}
 (current bounding box.south west)
  rectangle (current bounding box.north east);
\end{pgfonlayer}
\end{tikzpicture}

\vspace{1mm}
\pgfplotslegendfromname{groverlegend}

\caption{
Grover results across the evaluated thresholds.
(a) Task-success rates for the 4-qubit tagged-state, 8-qubit 3-SAT,
and 30-qubit tagged-state instances; error bars show the reported
$95\%$ confidence intervals. (b) Empirical total-variation distance from
the saturated output distribution. Solid curves use normalized-Born
sampling; dashed curves use early selection.
}
\label{fig:grovercorr}
\end{figure*}

The results show the expected advantage of thresholding in an
amplitude-amplification circuit. For tagged-state Grover at $\mathcal{T}=0.5$,
$2078.2\pm49.3$ of the $4080$ canonical endpoint-interference
reactions are performed, a $49.06\%$ reduction with $95\%$ CI
$[48.72,49.41]$. The empirical output distribution has TVD $0.379$
from the exact distribution. At
$\mathcal{T}=0.9$, the interference reduction is $50.38\%$ and
the empirical TVD is $0.324$. For tagged-state Grover, the marked state is returned in
$78\%$ of the reported trials at $\threshold=0.7$ and $94\%$ at
$\threshold=0.9$. At $\threshold=0.7$,
Table~\ref{tab:grover-results} reports a $62.07\%$ reduction in
canonical endpoint-interference reactions. Even at low thresholds,
the marked state is sampled more often than a uniformly random basis
state would be. For 3-SAT Grover at $\mathcal{T}=0.3$,
normalized-Born sampling performs $38.91\%$ of the canonical
interference reactions, omitting $61.09\%$. The empirical TVD is
$0.322$. Across the higher tested thresholds, the reported task-success estimates increase,
while endpoint-interference reduction and TVD are nonmonotonic.

We also evaluate an \emph{early termination} variant of the simulation. Rather than sampling from all molecules present when the simulator
halts, the early-selection variant returns the first endpoint molecule
that crosses the threshold. For the tagged-state instance, the marked
endpoint reaches amplitude about $0.68$, whereas incorrect endpoints
remain below about $0.185$. A threshold set above this
incorrect-endpoint ceiling can therefore identify the marked state
without waiting for full saturation, since any molecule crossing it
must correspond to the correct answer. 
For tagged-state Grover, its observed task-success rate rises
from $34\%$ at $\mathcal{T}=0.1$ to $95\%$ at
$\mathcal{T}=0.2$, $96\%$ at $\mathcal{T}=0.3$, and
$100\%$ for $\mathcal{T}\geq0.4$. For the 3-SAT instance it
rises from $34\%$ at $\mathcal{T}=0.1$ to $100\%$ observed
task success for $\mathcal{T}\geq0.2$. This is precisely the
regime in which endpoint-amplitude separation is useful: once
an incorrect endpoint can no longer be the first to cross the
threshold, the crossing endpoint provides a highly effective answer selector.

\begin{figure}[htb]
    \centering

    \subcaptionbox{}[0.48\textwidth]{%
        \centering
        \includegraphics[width=0.8\linewidth]{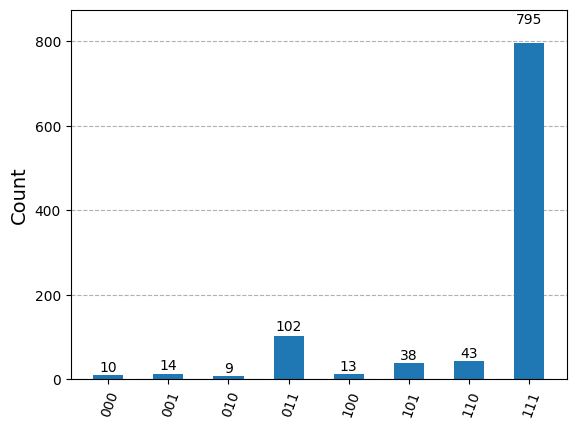}%
        \label{fig:groverfullit}%
    }\hfill
    \subcaptionbox{}[0.48\textwidth]{%
        \centering
        \includegraphics[width=0.8\linewidth]{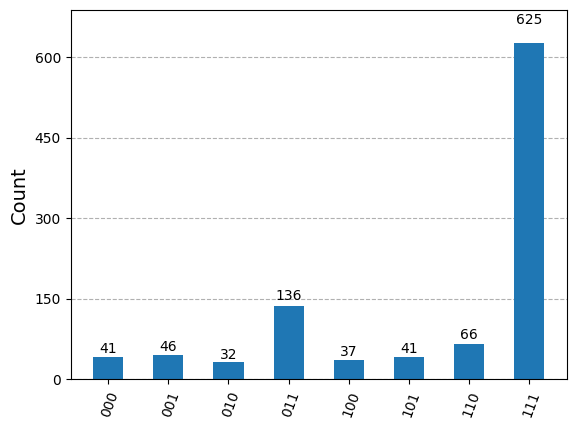}%
        \label{fig:grovlessit}%
    }

    \caption{Sampling counts for Grover’s algorithm with tagged state $\ket{111}$ on the IBM Brisbane machine: 
    (a) two iterations and 
    (b) a single iteration.}
    \label{fig:groveribmlessit}
  \end{figure}

A natural follow-up question is whether thresholding can be
combined with fewer Grover iterations. Each iteration increases the amplitude separation between marked and unmarked endpoints, but it
also increases the number of classically generated path
contributions. If early termination can discard enough
of the unmarked contributions, a circuit with fewer iterations may still produce a useful output distribution
at lower classical cost. Figure~\ref{fig:groveribmlessit}
shows IBM Brisbane sampling counts for a tagged $\ket{111}$ instance
with one and two iterations; these histograms provide qualitative
context for the changing marked-state prominence. Even with
a single iteration the marked state dominates the distribution, confirming that meaningful amplitude separation
exists before full amplification is complete. In the corresponding ChAM simulation, thresholds near 0.5 retain
more than 95\% marked-output accuracy while omitting roughly half of the endpoint-interference calculations. At this threshold, one full Grover iteration can be skipped
while preserving 95\% accuracy, a meaningful reduction in both quantum circuit depth and classical simulation
cost.

\begin{figure}[htb]
\centering
\begin{tikzpicture}
\begin{axis}[
    xlabel={\threshold},
    ylabel={Correct outputs (\%)},
    xmin=0, xmax=1,
    ymin=0, ymax=100,
    xtick={0.1,0.2,0.3,0.4,0.5,0.6,0.7,0.8,0.9},
    ytick={0,10,...,100},
    legend style={
        at={(0.02,0.98)},
        anchor=north west
    }
]
\addplot[color=cbBlue, mark=*, solid] coordinates {
    (0.1,8) (0.2,20) (0.3,55) (0.4,60)
    (0.5,64) (0.6,62) (0.7,65) (0.8,62) (0.9,61)
};
\addlegendentry{1 iteration}

\addplot[color=cbOrange, mark=square*, solid] coordinates {
    (0.1,5) (0.2,27) (0.3,59) (0.4,65)
    (0.5,69) (0.6,72) (0.7,70) (0.8,77) (0.9,78)
};
\addlegendentry{2 iterations}

\addplot[color=cbGreen, mark=triangle*, solid] coordinates {
    (0.1,20) (0.2,33) (0.3,48) (0.4,58)
    (0.5,69) (0.6,75) (0.7,79) (0.8,82) (0.9,92)
};
\addlegendentry{3 iterations}
\end{axis}
\end{tikzpicture}
\caption{\label{fig:grover5qubitlessit} Percentage of runs in which partial accumulation with threshold $\threshold$ produced the correct result under one, two, and three reduced iterations of Grover's algorithm.}
\end{figure}

\begin{figure*}[htb]
    \centering

    \subcaptionbox{}[0.48\textwidth]{%
        \centering
        \includegraphics[width=0.8\linewidth]{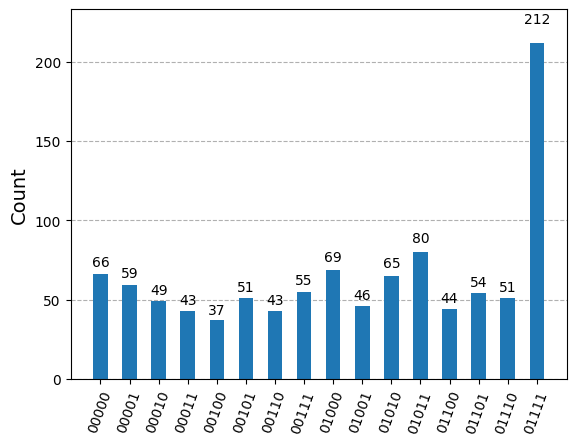}%
        \label{fig:g5lit}%
    }\hfill
    \subcaptionbox{}[0.48\textwidth]{%
        \centering
        \includegraphics[width=0.8\linewidth]{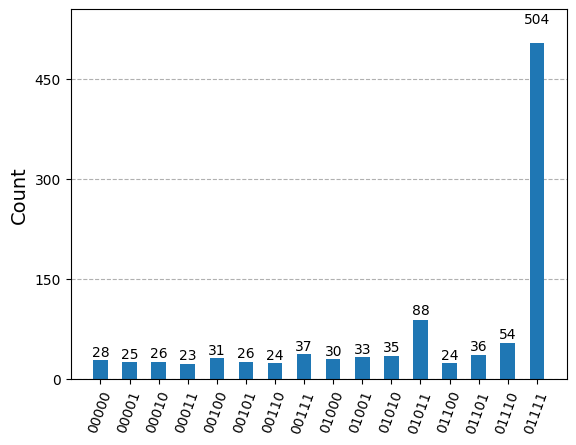}%
        \label{fig:g52it}%
    }

    \vspace{1em}

    \subcaptionbox{}[0.48\textwidth]{%
        \centering
        \includegraphics[width=0.8\linewidth]{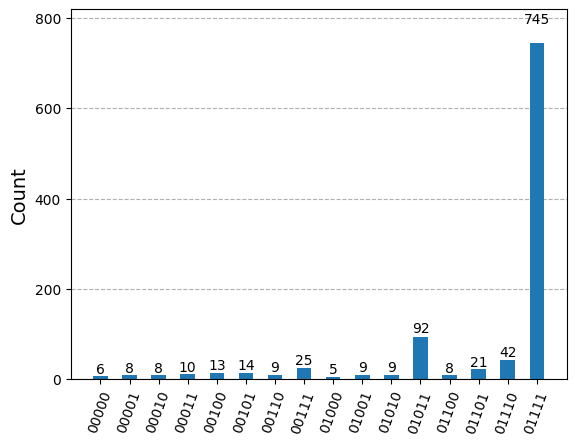}%
        \label{fig:git53it}%
    }\hfill
    \subcaptionbox{}[0.48\textwidth]{%
        \centering
        \includegraphics[width=0.8\linewidth]{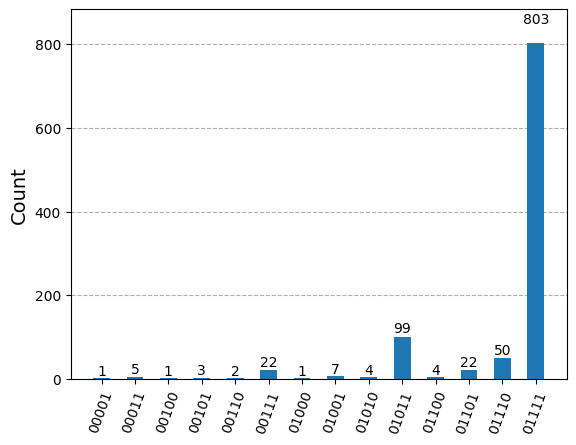}%
        \label{fig:g54it}%
    }

    \caption{Sampling counts for Grover’s algorithm with tagged state $\ket{01111}$ on the IBM Brisbane machine: 
    (a) one iteration, 
    (b) two iterations, 
    (c) three iterations, 
    and (d) four (optimal) iterations.}
    \label{fig:grov5qusearchits}
\end{figure*}

We examine this iteration-reduction
tradeoff in more detail using a 5-qubit Grover search with 3 ancilla
qubits (8 qubits total) and marked state $\ket{01111}$. The optimal
number of iterations for this instance is $\left\lfloor
\pi\sqrt{2^5}/4 \right\rfloor=4$. Figure~\ref{fig:grover5qubitlessit} shows the 
thresholded ChAM results for one through three iterations and Figure~\ref{fig:grov5qusearchits}
shows IBM Brisbane corresponding sampling counts for one through four iterations.
With one iteration the marked endpoint is not yet well separated from the rest: the 1024 correct intermediate states carry amplitudes of only $\pm 0.0056$, a small gap relative to the background. In this case, the thresholded simulation saturates near 60\% accuracy. With two iterations the amplitude separation grows and accuracy
rises to about 83\%, while still omitting a nontrivial fraction of endpoint reactions. With three iterations, one below the textbook optimum, accuracy reaches roughly
92\%. Taken together, these results support the interpretation that thresholding is most valuable when the circuit has already produced meaningful amplitude separation but full amplification has not yet been completed: the threshold exploits the gap that exists rather than waiting for the gap to be maximized.

\subsection{Simulating Simon and Shor Algorithms}
\label{sec:shor}

Simon and Shor circuits test a different regime from Grover search.
They do not amplify a single marked endpoint to the same degree;
instead, the useful information is distributed over a set of valid
outcomes. In a single Simon run, each valid endpoint receives only a
small number of terminal contributions. As $n$ grows, the maximum
amplitude available to any one endpoint decreases, making high
thresholds difficult to reach in a single execution.

We therefore run multiple independent circuit instances concurrently and aggregate them in one shared ChAM solution. This mirrors the textbook use of repeated Simon samples, but replaces sequential measurement and post-selection by a single joint endpoint distribution. Running $k$ instances simultaneously means each output state can accumulate contributions from all $k$ copies before the threshold is checked, boosting the maximum reachable amplitude and increasing the number of interference reactions each state can participate in. Concretely, after $k$ concurrent instances each state can participate in up to $2^{k-1}$ reactions, reaching a maximum amplitude of $k/2^{k-1}$. This is enough to make previously unreachable thresholds accessible, and we use the same strategy for the Shor period-finding experiments. 

We apply the task-success criteria defined in the experimental
protocol: membership in the valid-output set for Simon's algorithm
and successful recovery of non-trivial factors after classical
post-processing for Shor's algorithm.

We test two Simon instances corresponding to 6- and
8-qubit circuits, and two Shor instances factoring 15
and 21. Figures~\ref{fig:shor}
and~\ref{fig:shor3} use solid curves for normalized-Born sampling
and dashed curves for the early-selection variant, as defined in
Section~\ref{sec:experimental-protocol}.
Table~\ref{tab:simon-shor-work} reports the amount of omitted endpoint work
across both algorithm families.

\begin{figure*}[htb]
\centering

\resizebox{\linewidth}{!}{%
\begin{tikzpicture}

\begin{groupplot}[
    group style={
        group size=2 by 1,
        horizontal sep=1.5cm
    },
    width=0.47\linewidth,
    height=0.37\linewidth,
    xmin=0.08,
    xmax=0.92,
    xtick={0.1,0.2,0.3,0.4,0.5,0.6,0.7,0.8,0.9},
    xlabel={Threshold $\mathcal{T}$},
    grid=major,
    tick label style={font=\scriptsize},
    label style={font=\small},
    title style={font=\small}
]


\nextgroupplot[
    title={(a) Valid-output rate},
    ylabel={Valid outputs (\%)},
    ymin=0,
    ymax=105,
    ytick={0,20,40,60,80,100},
    legend to name=simonlegend,
    legend columns=2,
    legend style={
        font=\scriptsize,
        draw=none,
        /tikz/every even column/.append style={
            column sep=1.5em
        },
        /tikz/every odd column/.append style={
            column sep=0.4em
        }
    }
]


\addplot+[
    color=cbBlue,
    mark=*,
    solid,
    thick,
    error bars/.cd,
        y dir=both,
        y explicit
] table[
    x=T,
    y=mean,
    y error minus=emin,
    y error plus=eplus,
    row sep=crcr
] {
T   mean   emin   eplus \\
0.1 50     13.36  13.36 \\
0.2 50     13.36  13.36 \\
0.3 60     13.82  12.39 \\
0.4 70     13.75  10.90 \\
0.5 80     13.04   8.76 \\
0.6 78     13.24   9.25 \\
0.7 76     13.41   9.70 \\
0.8 80     13.04   8.76 \\
0.9 86     12.19   7.05 \\
};
\addlegendentry{$n=3$, Born}


\addplot+[
    color=cbRed,
    mark=triangle*,
    dashed,
    thick,
    error bars/.cd,
        y dir=both,
        y explicit
] table[
    x=T,
    y=mean,
    y error minus=emin,
    y error plus=eplus,
    row sep=crcr
] {
T   mean   emin   eplus \\
0.1 48     13.20  13.49 \\
0.2 72     13.67  10.53 \\
0.3 100     7.13   0 \\
0.4 100     7.13   0 \\
0.5 100     7.13   0 \\
0.7 100     7.13   0 \\
0.9 100     7.13   0 \\
};
\addlegendentry{$n=3$, early selection}


\addplot+[
    color=cbOrange,
    mark=square*,
    solid,
    thick,
    error bars/.cd,
        y dir=both,
        y explicit
] table[
    x=T,
    y=mean,
    y error minus=emin,
    y error plus=eplus,
    row sep=crcr
] {
T   mean   emin   eplus \\
0.1 56     13.69  12.84 \\
0.2 44     12.84  13.69 \\
0.3 50     13.36  13.36 \\
0.4 48     13.20  13.49 \\
0.5 52     13.49  13.20 \\
0.6 60     13.77  12.62 \\
0.7 70     13.03  13.60 \\
0.8 83     13.82  12.39 \\
0.9 76     13.81  11.24 \\
};
\addlegendentry{$n=5$, Born}


\addplot+[
    color=black,
    mark=square*,
    dashed,
    thick,
    error bars/.cd,
        y dir=both,
        y explicit
] table[
    x=T,
    y=mean,
    y error minus=emin,
    y error plus=eplus,
    row sep=crcr
] {
T   mean   emin   eplus \\
0.1 24      9.70  13.41 \\
0.2 36      7.66  12.51 \\
0.3 60     13.82  12.39 \\
0.4 50     13.03  13.60 \\
0.5 100     7.13   0 \\
0.6 100     7.13   0 \\
0.7 100     7.13   0 \\
0.8 100     7.13   0 \\
0.9 100     7.13   0 \\
};
\addlegendentry{$n=5$, early selection}


\nextgroupplot[
    title={(b) Distributional error},
    ylabel={Total-variation distance},
    ymin=0,
    ymax=1,
    ytick={0,0.2,0.4,0.6,0.8,1.0}
]


\addplot+[
    color=cbBlue,
    mark=*,
    solid,
    thick
] coordinates {
    (0.1,0.50)
    (0.2,0.50)
    (0.3,0.40)
    (0.4,0.30)
    (0.5,0.29)
    (0.6,0.23)
    (0.7,0.24)
    (0.8,0.23)
    (0.9,0.19)
};


\addplot+[
    color=cbRed,
    mark=triangle*,
    dashed,
    thick
] coordinates {
    (0.1,0.53)
    (0.2,0.28)
    (0.3,0.13)
    (0.4,0.18)
    (0.5,0.07)
    (0.7,0.22)
    (0.9, 0.22)
};


\addplot+[
    color=cbOrange,
    mark=square*,
    solid,
    thick
] coordinates {
    (0.1,0.533)
    (0.2,0.560)
    (0.3,0.517)
    (0.4,0.520)
    (0.5,0.497)   
    (0.6,0.455)
    (0.7,0.575)
    (0.8,0.493)
    (0.9,0.435)
};


\addplot+[
    color=black,
    mark=square*,
    dashed,
    thick
] coordinates {
    (0.1,0.875)
    (0.2,0.898)
    (0.3,0.875)
    (0.4,0.875)
    (0.5,0.875)
    (0.6,0.878)
    (0.7,0.938)
    (0.8,0.938)
    (0.9,0.938)
};

\end{groupplot}


\node[
    anchor=north,
    inner sep=0pt
] at
(
    $(group c1r1.south)!0.5!(group c2r1.south)
    +(0,-1.35cm)$
)
{\pgfplotslegendfromname{simonlegend}};

\begin{pgfonlayer}{background}
(current bounding box.south west)
  rectangle (current bounding box.north east);
\end{pgfonlayer}

\end{tikzpicture}%
}

\vspace{0.35cm}

\caption{
Simon results across the available threshold batches. Panel (a)
reports valid-output rate with the reported confidence intervals;
panel (b) reports empirical TVD from the exact saturated output
distribution. Solid curves use normalized-Born sampling and dashed
curves use early selection.
}
\label{fig:shor}

\end{figure*}

\begin{figure*}[htb]
\centering
\begin{tikzpicture}
\begin{groupplot}[
    group style={
        group size=2 by 1,
        horizontal sep=1.35cm
    },
    width=0.46\textwidth,
    height=0.34\textwidth,
    xmin=0.08,
    xmax=0.92,
    xtick={0.1,0.2,0.3,0.4,0.5,0.6,0.7,0.8,0.9},
    xlabel={Threshold $\mathcal{T}$},
    grid=major,
    tick label style={font=\scriptsize},
    label style={font=\small},
    title style={font=\small}
]

\nextgroupplot[
    title={(a) Shor task success},
    ylabel={Successful factor recovery (\%)},
    ymin=0,
    ymax=105,
    ytick={0,20,40,60,80,100},
    legend to name=shorlegend,
    legend columns=2,
    legend style={
        font=\scriptsize,
        draw=none,
        /tikz/every even column/.append style={column sep=1.2em},
        /tikz/every odd column/.append style={column sep=0.4em}
    }
]

\addplot+[
    black, solid, mark=*,
    error bars/.cd, y dir=both, y explicit
] table[x=T,y=mean,y error minus=emin,y error plus=eplus] {
T mean emin eplus
0.1 36 11.86 13.86
0.2 42 12.62 13.77
0.3 46 13.03 13.60
0.4 48 13.20 13.49
0.5 62 13.85 12.14
0.6 56 13.49 13.20
0.7 64 13.60 13.03
0.8 72 13.77 12.62
0.9 80 13.82 12.39
};
\addlegendentry{$N=15$, Born}

\addplot+[
    black, dashed, mark=square*,
    error bars/.cd, y dir=both, y explicit
] table[x=T,y=mean,y error minus=emin,y error plus=eplus] {
T mean emin eplus
0.1 30 10.90 13.75
0.2 56 13.69 12.84
0.3 40 12.39 13.82
0.4 42 12.62 13.77
0.5 52 13.49 13.20
0.6 98 8.50 1.65
0.7 100 7.13 0
0.8 100 7.13 0
0.9 100 7.13 0
};
\addlegendentry{$N=15$, early selection}

\addplot+[
    orange, solid, mark=triangle*,
    error bars/.cd, y dir=both, y explicit
] table[x=T,y=mean,y error minus=emin,y error plus=eplus] {
T mean emin eplus
0.1 32 11.24 13.81
0.2 36 11.86 13.86
0.3 40 12.39 13.82
0.4 38 12.14 13.85
0.5 51 13.42 13.28
0.6 61 12.50 13.80
0.7 55 12.94 13.65
0.8 61 12.50 13.80
0.9 64 13.60 13.03
};
\addlegendentry{$N=21$, Born}

\addplot+[
    orange, dashed, mark=diamond*,
    error bars/.cd, y dir=both, y explicit
] table[x=T,y=mean,y error minus=emin,y error plus=eplus] {
T mean emin eplus
0.1 30 10.90 13.75
0.2 57 13.73 12.73
0.3 43 12.73 13.73
0.4 46 13.03 13.60
0.5 56 13.69 12.84
0.6 100 7.13 0
0.7 98 8.50 1.65
0.8 98 10.22 3.94
0.9 100 7.13 0
};
\addlegendentry{$N=21$, early selection}

\nextgroupplot[
    title={(b) Distributional error},
    ylabel={Total-variation distance},
    ymin=0,
    ymax=1,
    ytick={0,0.2,0.4,0.6,0.8,1.0}
]

\addplot+[black, solid, mark=*] coordinates {
(0.1,0.82) (0.2,0.82) (0.3,0.78)
(0.4,0.76) (0.5,0.64) (0.6,0.72)
(0.7,0.72) (0.8,0.60) (0.9,0.58)
};

\addplot+[black, dashed, mark=square*] coordinates {
(0.1,0.86) (0.2,0.80) (0.3,0.78)
(0.4,0.80) (0.5,0.74) (0.6,0.69)
(0.7,0.69) (0.8,0.69) (0.9,0.65)
};

\addplot+[orange, solid, mark=triangle*] coordinates {
(0.1,0.84) (0.2,0.85) (0.3,0.81)
(0.4,0.80) (0.5,0.69) (0.6,0.77)
(0.7,0.76) (0.8,0.68) (0.9,0.61)
};

\addplot+[orange, dashed, mark=diamond*] coordinates {
(0.1,0.86) (0.2,0.79) (0.3,0.76)
(0.4,0.78) (0.5,0.72) (0.6,0.68)
(0.7,0.70) (0.8,0.72) (0.9,0.63)
};

\end{groupplot}

\node[
    anchor=north,
    inner sep=0pt
] at
($(group c1r1.south)!0.5!(group c2r1.south)+(0,-0.9cm)$)
{\pgfplotslegendfromname{shorlegend}};

\begin{pgfonlayer}{background}
(current bounding box.south west)
  rectangle (current bounding box.north east);
\end{pgfonlayer}
\end{tikzpicture}
\caption{
Shor results for $N=15$ and $N=21$. Panel (a) reports the
fraction of trials for which classical post-processing recovers
non-trivial factors of the corresponding $N$, with the reported
confidence intervals. Panel (b) reports empirical TVD from the
corresponding saturated phase distribution. 
}
\label{fig:shor3}
\end{figure*}

\begin{table*}[htb]
\centering
\caption{
Simon and Shor results for normalized-Born (Born) and
early-selection (Early) output. $B_{\mathrm{active}}$ and
$I_{\mathcal T}$ are reported as mean $\pm$ sample standard
deviation. Reduction and task success are reported with $95\%$
confidence intervals; TVD is reported with its $95\%$ bootstrap
confidence interval.
}
\label{tab:simon-shor-work}

\scriptsize
\setlength{\tabcolsep}{2.5pt}
\renewcommand{\arraystretch}{1.08}

\resizebox{\dimexpr\textwidth-2\fboxsep\relax}{!}{%
\begin{tabular}{lllrrrrrr}
\toprule
Instance &
Policy &
$\mathcal{T}$ &
$B_{\mathrm{active}}$ &
$B_c/B$ &
$I_{\mathcal T}/I_{\mathrm{full}}$ &
Reduction \% [95\% CI] &
Success \% [95\% CI] &
TVD [95\% CI] \\
\midrule


\multirow{10}{*}{Simon $n=3$}

& \multirow{5}{*}{Born}
& 0.1
& $1.0\pm0.0$
& $24/192$
& $0.0\pm0.0\,/\,40$
& 100.00 [100.00,100.00]
& 50 [36.6,63.4]
& 0.500 [0.380,0.640] \\

&
& 0.3
& $19.7\pm2.3$
& $24/192$
& $1.3\pm0.5\,/\,40$
& 96.65 [96.28,97.02]
& 60 [46.2,72.4]
& 0.400 [0.280,0.540] \\

&
& 0.5
& $20.2\pm2.4$
& $24/192$
& $8.6\pm1.7\,/\,40$
& 78.60 [77.36,79.84]
& 80 [67.0,88.8]
& 0.290 [0.170,0.430] \\

&
& 0.7
& $21.6\pm2.1$
& $24/192$
& $8.8\pm1.4\,/\,40$
& 78.00 [77.01,78.99]
& 76 [62.6,85.7]
& 0.240 [0.160,0.380] \\

&
& 0.9
& $24.0\pm3.2$
& $24/192$
& $14.2\pm2.1\,/\,40$
& 64.50 [62.98,66.02]
& 86 [73.8,93.0]
& 0.190 [0.110,0.330] \\

\cmidrule(lr){2-9}

& \multirow{5}{*}{Early}
& 0.1
& $1.0\pm0.0$
& $24/192$
& $0.0\pm0.0\,/\,40$
& 100.00 [100.00,100.00]
& 48 [34.8,61.5]
& 0.530 [0.440,0.660] \\

&
& 0.3
& $17.9\pm3.9$
& $24/192$
& $6.7\pm2.1\,/\,40$
& 83.30 [81.79,84.81]
& 100 [92.9,100.0]
& 0.130 [0.060,0.270] \\

&
& 0.5
& $31.4\pm2.2$
& $24/192$
& $21.0\pm2.3\,/\,40$
& 47.45 [45.80,49.10]
& 100 [92.9,100.0]
& 0.070 [0.040,0.220] \\

&
& 0.7
& $40.1\pm2.4$
& $24/192$
& $28.2\pm2.1\,/\,40$
& 29.55 [28.05,31.05]
& 100 [92.9,100.0]
& 0.220 [0.110,0.340] \\

&
& 0.9
& $133.7\pm5.4$
& $24/192$
& $18.2\pm2.1\,/\,40$
& 54.50 [50.05,61.28]
& 100 [92.9,100.0]
& 0.220 [0.110,0.340] \\

\midrule


\multirow{10}{*}{Simon $n=5$}

& \multirow{5}{*}{Born}
& 0.1
& $1.0\pm0.0$
& $160/5120$
& $0.0\pm0.0\,/\,288$
& 100.00 [100.00,100.00]
& 56 [42.3,68.8]
& 0.533 [0.510,0.690] \\

&
& 0.3
& $15.3\pm1.6$
& $160/5120$
& $7.5\pm0.8\,/\,288$
& 97.40 [97.32,97.49]
& 50 [36.6,63.4]
& 0.517 [0.458,0.678] \\

&
& 0.5
& $129.5\pm1.8$
& $160/5120$
& $97.8\pm1.7\,/\,288$
& 66.03 [65.86,66.20]
& 52 [38.5,65.2]
& 0.498 [0.450,0.658] \\

&
& 0.7
& $131.7\pm1.3$
& $160/5120$
& $100.2\pm1.4\,/\,288$
& 65.22 [65.07,65.36]
& 70 [56.97, 83.60]
& 0.575 [0.515,0.733] \\

&
& 0.9
& $139.8\pm3.3$
& $160/5120$
& $107.3\pm3.2\,/\,288$
& 62.74 [62.42,63.05]
& 76 [62.19, 87.24]
& 0.435 [0.393,0.595] \\

\cmidrule(lr){2-9}

& \multirow{5}{*}{Early}
& 0.1
& $1.0\pm0.0$
& $160/5120$
& $0.0\pm0.0\,/\,288$
& 100.00 [100.00,100.00]
& 24 [14.3,37.4]
& 0.875 [0.875,0.918] \\

&
& 0.3
& $97.5\pm1.7$
& $160/5120$
& $7.5\pm1.4\,/\,288$
& 97.39 [97.25,97.53]
& 60 [46.2,72.4]
& 0.875 [0.875,0.875] \\

&
& 0.5
& $320.5\pm1.4$
& $160/5120$
& $97.4\pm1.4\,/\,288$
& 66.19 [66.05,66.33]
& 100 [92.9,100.0]
& 0.875 [0.875,0.918] \\

&
& 0.7
& $1598.7\pm1.4$
& $160/5120$
& $99.6\pm1.4\,/\,288$
& 65.42 [65.29,65.56]
& 100 [92.9,100.0]
& 0.938 [0.938,0.938] \\

&
& 0.9
& $3223.1\pm2.7$
& $160/5120$
& $105.0\pm2.7\,/\,288$
& 63.55 [63.29,63.81]
& 100 [92.9,100.0]
& 0.938 [0.938,0.938] \\

\midrule


\multirow{10}{*}{Shor $N=15$}

& \multirow{5}{*}{Born}
& 0.1
& $204.6\pm6.4$
& $1120/2560$
& $9.35\pm0.2\,/\,2496$
& 99.63 [99.55,99.96]
& 36 [24.1,49.9]
& 0.820 [0.720,0.920] \\

&
& 0.3
& $303.5\pm7.9$
& $1120/2560$
& $171.66\pm4.7\,/\,2496$
&  93.12 [93.07,93.17]
& 46 [33.0,59.6]
& 0.780 [0.660,0.880] \\

&
& 0.5
& $800.2\pm12.1$
& $1120/2560$
& $630.9\pm11.4\,/\,2496$
& 74.72 [74.60,74.85]
& 62 [48.2,74.1]
& 0.640 [0.530,0.760] \\

&
& 0.7
& $1142.4\pm15.4$
& $1120/2560$
& $1258.4\pm15.5\,/\,2496$
& 49.58 [49.41,49.76]
& 64 [50.4, 77.03]
& 0.720 [0.600,0.840] \\

&
& 0.9
& $1184.4\pm15.4$
& $1120/2560$
& $1673.6\pm15.0\,/\,2496$
& 32.95 [32.78,33.12]
& 80 [66.18,92.39]
& 0.580 [0.460,0.720] \\

\cmidrule(lr){2-9}

& \multirow{5}{*}{Early}
& 0.1
& $230.8\pm5.5$
& $1120/2560$
& $12\pm0.5\,/\,2496$
&  99.52 [99.51,99.52]
& 30 [19.1,43.8]
& 0.860 [0.760,0.940] \\

&
& 0.3
& $646.7\pm24.5$
& $1120/2560$
& $342.04\pm12.4\,/\,2496$
& 86.30 [86.16,86.43]
& 40 [27.6,53.8]
& 0.780 [0.750,0.880] \\

&
& 0.5
& $1178.6\pm10.6$
& $1120/2560$
& $943.5\pm9.9\,/\,2496$
& 62.20 [62.09,62.31]
& 52 [38.5,65.2]
& 0.740 [0.670,0.860] \\

&
& 0.7
& $1941.1\pm12.4$
& $1120/2560$
& $1711.35\pm12.3\,/\,2496$
& 31.44 [31.30,31.57]
& 100 [92.9,100.0]
& 0.690 [0.610,0.750] \\

&
& 0.9
& $2204.9\pm13.9$
& $1120/2560$
& $1168.0\pm13.8\,/\,2496$
& 53.21 [53.05,53.36]
& 100 [92.9,100.0]
& 0.650 [0.570,0.730] \\

\midrule


\multirow{10}{*}{Shor $N=21$}

& \multirow{5}{*}{Born}
& 0.1
& $21{,}456\pm5{,}582$
& $10{,}240/10{,}485{,}760$
& $959.4\pm174.4\,/\,10{,}223{,}616$
& 99.99 [99.99,99.99]
& 32 [20.8,45.8]
& 0.840 [0.740,0.940] \\

&
& 0.3
& $303{,}222\pm71{,}506$
& $10{,}240/10{,}485{,}760$
& $169{,}261\pm42{,}542\,/\,10{,}223{,}616$
& 98.34 [98.23,98.46]
& 40 [27.6,53.8]
& 0.810 [0.690,0.910] \\

&
& 0.5
& $4{,}145{,}728\pm625{,}478$
& $10{,}240/10{,}485{,}760$
& $3{,}303{,}142\pm589{,}293\,/\,10{,}223{,}616$
& 67.69 [66.09,69.29]
& 51 [37.6,64.3]
& 0.690 [0.580,0.810] \\

&
& 0.7
& $7{,}899{,}316\pm854{,}280$
& $10{,}240/10{,}485{,}760$
& $6{,}956{,}280\pm859{,}827\,/\,10{,}223{,}616$
& 31.96 [29.63,34.29]
& 55 [42.06,68.65]
& 0.760 [0.640,0.880] \\

&
& 0.9
& $9{,}677{,}312\pm808{,}192$
& $10{,}240/10{,}485{,}760$
& $8{,}795{,}648\pm787{,}200\,/\,10{,}223{,}616$
& 13.97 [11.83,16.10]
& 64 [50.4, 77.03]
& 0.610 [0.490,0.750] \\

\cmidrule(lr){2-9}

& \multirow{5}{*}{Early}
& 0.1
& $20{,}758\pm4{,}797$
& $10{,}240/10{,}485{,}760$
& $1{,}046.6\pm436.1\,/\,10{,}223{,}616$
& 99.99 [99.99,99.99]
& 30 [19.1,43.8]
& 0.860 [0.760,0.940] \\

&
& 0.3
& $422{,}701\pm221{,}759$
& $10{,}240/10{,}485{,}760$
& $223{,}570\pm112{,}237\,/\,10{,}223{,}616$
& 97.81 [97.51,98.12]
& 43 [30.3,56.7]
& 0.760 [0.730,0.860] \\

&
& 0.5
& $4{,}063{,}020\pm547{,}939$
& $10{,}240/10{,}485{,}760$
& $3{,}230{,}773\pm511{,}754\,/\,10{,}223{,}616$
& 68.40 [67.01,69.79]
& 56 [42.3,68.8]
& 0.720 [0.650,0.840] \\

&
& 0.7
& $7{,}827{,}201\pm687{,}862$
& $10{,}240/10{,}485{,}760$
& $6{,}900{,}807\pm682{,}315\,/\,10{,}223{,}616$
& 32.50 [30.65,34.35]
& 98 [89.5,99.6]
& 0.700 [0.620,0.760] \\

&
& 0.9
& $9{,}703{,}552\pm729{,}472$
& $10{,}240/10{,}485{,}760$
& $8{,}816{,}640\pm724{,}224\,/\,10{,}223{,}616$
& 13.76 [11.80,15.73]
& 100 [92.9,100.0]
& 0.630 [0.550,0.710] \\

\bottomrule
\end{tabular}
}

\end{table*}

The Simon data show that moderate thresholds retain useful accuracy
while omitting substantial work. Under
normalized-Born sampling, the $n=3$ instance improves from a
valid-output rate of $50\%$ at $\mathcal{T}=0.1$ to
$86\%$ at $\mathcal{T}=0.9$ (Wilson 95\% CI
$[73.8,93.0]$). At the latter threshold only
$14.2\pm2.1$ of the 40 canonical endpoint-interference
reactions have been performed, corresponding to a
$64.50\%$ reduction, while the empirical TVD has fallen to
$0.19$. At $\threshold=0.7$, the $n=3$
instance returns valid outcomes in 76\% of runs, and the
early-selection variant reaches 95\%; the $n=5$ instance reaches 70\%
in the base simulation and 90\% with early selection while omitting
roughly 1900 terminal contributions. The improvement from early
selection is more pronounced here than in Grover search because the
valid endpoints in Simon are not uniquely amplified: multiple
endpoints carry comparable amplitudes so returning the first one
to cross the threshold is a reliable strategy as long as the threshold
is calibrated above the noise floor. Early termination selection is particularly effective for the
$n=3$ instance. At $\mathcal{T}=0.3$ all 50 completed trials
return a valid output while omitting $83.30\%$ of the
canonical endpoint reactions; at $\mathcal{T}=0.5$, all
50 trials again return valid outputs and the empirical TVD is
only $0.07$ while $47.45\%$ of endpoint reactions remain
unperformed. 

The Shor results are more conservative, as expected from the flatter
useful-output distribution. The Shor experiment evaluates 
complete classical success criterion following continued fraction order recovery rather than judging the raw phase sample alone. Each thresholded output is converted to a phase estimate, candidate orders are tested, and a run is classified as successful only when non-trivial factors are recovered.

Normalized-Born sampling exhibits substantial interference
reduction throughout the tested range. At
$\mathcal{T}=0.5$, only $63.9\pm11.4$ of the 2496
canonical endpoint-interference reactions have completed,
corresponding to a $97.44\%$ reduction; factor recovery
succeeds in $62\%$ of trials (Wilson 95\% CI
$[48.2,74.1]$). At $\mathcal{T}=0.9$, the reduction remains
$93.28\%$, factor recovery succeeds in $60\%$ of trials,
and empirical TVD is $0.58$.

\begin{figure}[htb]
    \centering
    \includegraphics[width=0.9\linewidth]{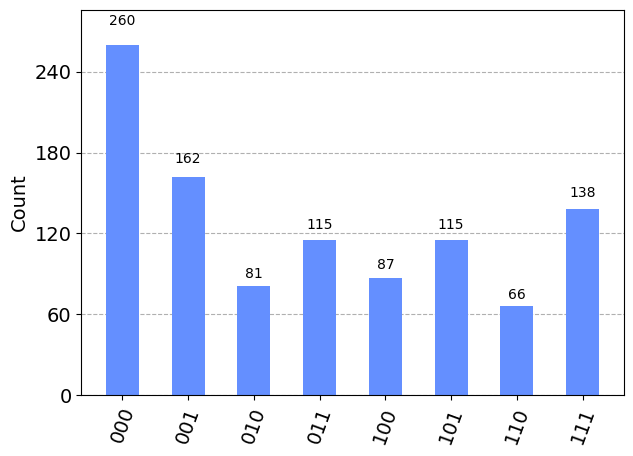}
\caption{\label{fig:shorquant}IBM Brisbane sampling counts for a
small $N=15$ Shor circuit.}
\end{figure}

Early termination selection again demonstrates the distinction
between task-oriented and distribution-preserving
approximation. Factor recovery rises to $98\%$ at
$\mathcal{T}=0.6$ and to $100\%$ in all 50 trials for
$\mathcal{T}\geq0.7$, while still omitting approximately
$95\%$ of canonical endpoint reactions at
$\mathcal{T}=0.7$. The corresponding TVD, however, remains
approximately $0.69$. Thus the early-selection rule can bias
the partial interference process toward phase estimates that
are especially useful to the classical post-processing stage
without faithfully reproducing the complete saturated phase
distribution. For $N=21$ and $\threshold=0.7$,
the simulation omits more than 2.5 million terminal contributions. When correct and incorrect
endpoints have comparable amplitudes, an incorrect endpoint can cross
the threshold first, and accuracy degrades at high thresholds rather
than improving. This is the main limitation of thresholded endpoint
selection. Figure~\ref{fig:shorquant} shows IBM Brisbane sampling counts for the
$N=15$ circuit. The broad phase distribution motivates the
separation between distributional fidelity and downstream task
success used above: a thresholded simulation need not
reproduce every phase with the correct frequency in order for
the subsequent classical order-recovery procedure to obtain
the correct factors.

\subsection{Empirical Amplitude Separation and Threshold Reliability}
\label{sec:empirical-separation}

Statistical interference sampling is built on multiple core observations. (1) A correct state has a higher probability of exceeding the threshold first (2) At the point in time where the threshold is exceeded, more incorrect states will have been computationally omitted compared to correct ones (3) For a sufficiently high threshold, we can guarantee enough correct states have contributed. In this section we measure and present the \textit{developedAmplitudeMagnitude}, the amplitude of the state that exceeds $\tau$ and triggers the membrane shutdown, and \textit{maxIncorrectReadyAmplitudeAtSampling} the maximum amplitude among all incorrect tagged states. In the case of \textit{early-termination}, \textit{developedAmplitudeMagnitude} equals the amplitude of the selected the state. We formalize these observations mathematically in Appendix  \ref{sec:formal} but present a summary of observed amplitudes in further support.

\begin{table*}[t]
\centering
\caption{
Representative early termination configurations around the
transition to reliable selection. $|A_{\mathrm{sel}}|$ is the
aggregate magnitude of the selected endpoint,
$A_{\mathrm{inc}}^{\max}$ is the largest incorrect
ready-state aggregate magnitude, and
$M_{\mathrm{sel}}
=|A_{\mathrm{sel}}|-A_{\mathrm{inc}}^{\max}$.
Amplitude quantities are mean $\pm$ sample standard deviation
over 50 executions. $R_{\mathcal T}$ is the percentage of
canonical endpoint-interference reactions omitted.
}
\label{tab:selection-separation}
\small
\setlength{\tabcolsep}{5pt}
\resizebox{\textwidth}{!}{%
\begin{tabular}{llrrrrr}
\toprule
Instance &
$\mathcal{T}$ &
$|A_{\mathrm{sel}}|$ &
$A_{\mathrm{inc}}^{\max}$ &
$M_{\mathrm{sel}}$ &
$R_{\mathcal T}$ (\%) &
Task success (\%) \\
\midrule

Tagged Grover
& 0.2
& $0.2034\pm0.0022$
& $0.2034\pm0.0022$
& $0.0000\pm0.0000$
& 96.17 & 0 \\

& 0.3
& $0.3122\pm0.0022$
& $0.2481\pm0.0301$
& $0.0641\pm0.0302$
& 70.57 & 96 \\

& 0.4
& $0.4069\pm0.0031$
& $0.2087\pm0.0129$
& $0.1981\pm0.0136$
& 55.53 & 100 \\

\midrule

3-SAT Grover
& 0.1
& $0.1051\pm0.0008$
& $0.1051\pm0.0008$
& $0.0000\pm0.0000$
& 98.47 & 0 \\

& 0.2
& $0.2044\pm0.0000$
& $0.1068\pm0.0033$
& $0.0976\pm0.0033$
& 75.12 & 100 \\

& 0.3
& $0.3042\pm0.0013$
& $0.1587\pm0.0025$
& $0.1455\pm0.0026$
& 61.09 & 100 \\

\midrule

Simon $n=3$
& 0.2
& $0.2625\pm0.0379$
& $0.1825\pm0.0629$
& $0.0800\pm0.0747$
& 96.20 & 72 \\

& 0.3
& $0.3850\pm0.0343$
& $0.2075\pm0.0598$
& $0.1775\pm0.0672$
& 83.30 & 100 \\

& 0.5
& $0.6250\pm0.0000$
& $0.1950\pm0.0627$
& $0.4300\pm0.0627$
& 47.45 & 100 \\

\midrule

Simon $n=5$
& 0.3
& $0.4325\pm0.0629$
& $0.5000\pm0.0000$
& $-0.0675\pm0.0629$
& 97.39 & 60 \\

& 0.4
& $0.5000\pm0.0000$
& $0.5000\pm0.0000$
& $0.0000\pm0.0000$
& 97.12 & 46 \\

& 0.5
& $0.6250\pm0.0000$
& $0.5000\pm0.0000$
& $0.1250\pm0.0000$
& 66.19 & 100 \\

\midrule

Shor $N=15$
& 0.5
& $0.5200\pm0.0256$
& $0.4900\pm0.0271$
& $0.0300\pm0.0441$
& 97.50 & 52 \\

& 0.6
& $0.6262\pm0.0088$
& $0.5144\pm0.0289$
& $0.1118\pm0.0308$
& 96.20 & 98 \\

& 0.7
& $0.7512\pm0.0088$
& $0.5508\pm0.0356$
& $0.2005\pm0.0353$
& 95.02 & 100 \\

\bottomrule
\end{tabular}
} 
\end{table*}

\begin{figure}[t]
\centering
\begin{tikzpicture}
\begin{axis}[
    width=\linewidth,
    height=0.62\linewidth,
    xlabel={Threshold $\mathcal{T}$},
    ylabel={Selection-separation margin $M_{\mathrm{sel}}$},
    xmin=0.08, xmax=0.92,
    ymin=-0.18, ymax=0.62,
    xtick={0.1,0.2,0.3,0.4,0.5,0.6,0.7,0.8,0.9},
    grid=major,
    legend style={
        at={(0.5,-0.28)},
        anchor=north,
        legend columns=2,
        font=\scriptsize,
        draw=none
    },
    tick label style={font=\scriptsize},
    label style={font=\small}
]

\addplot[
    black,
    densely dotted,
    domain=0.1:0.9
] {0};

\addplot+[
    solid,
    mark=*,
    error bars/.cd,
    y dir=both,
    y explicit
] table[
    x=T,
    y=margin,
    y error=sd
] {
T   margin    sd
0.1 0.00000 0.00000
0.2 0.00000 0.00000
0.3 0.06406 0.03015
0.4 0.19812 0.01356
0.5 0.29406 0.01832
0.6 0.26531 0.01984
0.7 0.10875 0.02823
0.8 0.39500 0.08352
0.9 0.52969 0.01105
};
\addlegendentry{Tagged Grover}

\addplot+[
    dashed,
    mark=square*,
    error bars/.cd,
    y dir=both,
    y explicit
] table[
    x=T,
    y=margin,
    y error=sd
] {
T   margin    sd
0.1 0.00000 0.00000
0.2 0.09756 0.00327
0.3 0.14551 0.00264
0.4 0.11358 0.00449
0.5 0.27235 0.01188
0.6 0.15125 0.00739
0.7 0.20451 0.00505
0.8 0.31908 0.00345
0.9 0.28715 0.00425
};
\addlegendentry{3-SAT Grover}

\addplot+[
    dashdotted,
    mark=triangle*,
    error bars/.cd,
    y dir=both,
    y explicit
] table[
    x=T,
    y=margin,
    y error=sd
] {
T   margin     sd
0.1  0.03000  0.05393
0.2 -0.00500  0.09091
0.3 -0.06750  0.06293
0.4  0.00000  0.00000
0.5  0.12500  0.00000
0.6  0.12750  0.01768
0.7  0.25500  0.02474
0.8  0.38000  0.02474
0.9  0.50000  0.00000
};
\addlegendentry{Simon $n=5$}

\addplot+[
    densely dashed,
    mark=diamond*,
    error bars/.cd,
    y dir=both,
    y explicit
] table[
    x=T,
    y=margin,
    y error=sd
] {
T   margin     sd
0.1 -0.02501 0.06567
0.2 -0.00398 0.06149
0.3 -0.01744 0.07477
0.4  0.03367 0.04917
0.5  0.03001 0.04414
0.6  0.11181 0.03083
0.7  0.20046 0.03527
0.8  0.23635 0.03800
0.9  0.29792 0.05048
};
\addlegendentry{Shor $N=15$}

\end{axis}
\end{tikzpicture}

\caption{
Empirical selection-separation margin
$M_{\mathrm{sel}}
 = |A_{\mathrm{sel}}|-A_{\mathrm{inc}}^{\max}$
for early termination selection. Error bars show one sample
standard deviation across 50 independent executions and
therefore represent run-to-run scheduling variability rather
than uncertainty in the mean. The dotted horizontal line
marks equal selected and maximum-incorrect amplitudes.
Large, consistently positive margins coincide with the
regimes in which early termination selection becomes reliable.}

\label{fig:selection-margin}
\end{figure}

The repeated experiments also allow us to examine directly the
amplitude-separation mechanism underlying early-termination
selection. Figure~\ref{fig:selection-margin} reports the
selection-separation margin
\[ \text{$\displaystyle 
  M_{\mathrm{sel}}
  =
  |A_{\mathrm{sel}}|
  -
  A_{\mathrm{inc}}^{\max} $}
\]
at the instant of selection. The error bars show one sample
standard deviation rather than a confidence interval for the
mean, since the purpose of this figure is to expose
run-to-run variation in the separation encountered by the
scheduler.

The results closely track the observed transition to reliable
early termination selection. For tagged-state Grover,
$M_{\mathrm{sel}}$ is zero at
$\mathcal{T}=0.1$ and $0.2$, where the marked endpoint is
never selected in the 50 trials. At $\mathcal{T}=0.3$ the
mean margin becomes $0.0641\pm0.0302$ and the marked state is
returned in 48 of 50 trials. At $\mathcal{T}=0.4$ the margin
increases to $0.1981\pm0.0136$, and all 50 trials select the
marked state. The 3-SAT Grover instance exhibits an even
sharper transition: at $\mathcal{T}=0.1$ the mean margin is
zero and none of the 50 trials returns the satisfying
assignment, whereas at $\mathcal{T}=0.2$ the margin is
$0.0976\pm0.0033$ and all 50 trials return it.

The same behavior appears in the more weakly separated
algorithm families, but at larger thresholds. For Simon
$n=5$, the mean margin is $-0.0675\pm0.0629$ at
$\mathcal{T}=0.3$ and approximately zero at
$\mathcal{T}=0.4$, where valid-output rates are $60\%$ and
$46\%$, respectively. At $\mathcal{T}=0.5$ the selected
endpoint is separated from the largest incorrect ready
endpoint by $0.125$, and all 50 trials return valid outputs.
For Shor $N=15$, small margins around zero at
$\mathcal{T}=0.4$--$0.5$ correspond to factor-recovery rates
of only $42\%$--$52\%$. When the margin increases to
$0.1118\pm0.0308$ at $\mathcal{T}=0.6$, factor recovery
increases to $98\%$; at $\mathcal{T}=0.7$, a margin of
$0.2005\pm0.0353$ coincides with successful factor recovery
in all 50 trials.

$M_{\mathrm{sel}}$ is measured from the
realized, scheduler-dependent ready population, and a mean
positive margin does not imply that the margin is positive in
every execution. They do, however, provide direct empirical
support for the amplitude-gap interpretation developed in
Appendix \ref{sec:formal}. Reliable early termination emerges when the selected
endpoint develops a margin that is large relative to the
run-to-run variation of the incorrect endpoint population.

Table~\ref{tab:threshold_task_success} reports the complete
threshold sweep rather than selecting only configurations near
the transition to high task success. The results show that a
positive mean separation margin is not by itself sufficient to
guarantee correct selection. For example, Simon $n=5$ has a
small positive mean margin at $\mathcal{T}=0.1$ while only
$24\%$ of trials return valid outputs. Likewise, because the
reported margin is averaged across independent executions, a
negative or near-zero mean does not imply that every individual
execution lacks separation.

A clearer pattern emerges once the selected endpoint develops
a margin that is large relative to the scheduler-induced
run-to-run variation. Tagged Grover transitions from zero mean
separation and zero task success at
$\mathcal{T}\leq0.2$ to
$M_{\mathrm{sel}}=0.064$ and $96\%$ success at
$\mathcal{T}=0.3$, followed by $100\%$ observed success at all
larger thresholds. Simon $n=5$ reaches $100\%$ observed task
success beginning at $\mathcal{T}=0.5$, where the mean margin
increases to $0.125$. Shor $N=15$ exhibits a similar but more
gradual transition: margins remain close to zero through
$\mathcal{T}=0.5$, where factor-recovery success is $52\%$,
then increase to $0.112$ at $\mathcal{T}=0.6$ and $0.200$ at
$\mathcal{T}=0.7$, where observed factor-recovery success rises
to $98\%$ and $100\%$, respectively.

These results therefore support an empirical association
between sufficiently developed endpoint-amplitude separation
and reliable early termination selection, rather than a
deterministic correspondence between the sign of
$M_{\mathrm{sel}}$ and correctness.

\begin{table}[t]
\centering
\caption{
Threshold-dependent empirical selection separation and
task success under early termination selection. The
selection-separation margin is
$M_{\mathrm{sel}}
=|A_{\mathrm{sel}}|-A_{\mathrm{inc}}^{\max}$.
For Shor's algorithm, task success denotes successful factor
recovery. Each entry is computed from 50 independent trials.
}
\label{tab:threshold_task_success}
\small
\setlength{\tabcolsep}{5pt}
\begin{tabular}{lccc}
\toprule
\textbf{Algorithm} &
$\mathcal{T}$ &
$M_{\mathrm{sel}}$ &
\textbf{Task success} \\
\midrule

\multirow{9}{*}{Tagged Grover}
    & 0.1 & 0.000 & 0\%   \\
    & 0.2 & 0.000 & 0\%   \\
    & 0.3 & 0.064 & 96\%  \\
    & 0.4 & 0.198 & 100\% \\
    & 0.5 & 0.294 & 100\% \\
    & 0.6 & 0.265 & 100\% \\
    & 0.7 & 0.109 & 100\% \\
    & 0.8 & 0.395 & 100\% \\
    & 0.9 & 0.530 & 100\% \\
\midrule

\multirow{9}{*}{Simon $n=5$}
    & 0.1 & 0.030  & 24\%  \\
    & 0.2 & $-0.005$ & 16\%  \\
    & 0.3 & $-0.068$ & 60\%  \\
    & 0.4 & 0.000  & 46\%  \\
    & 0.5 & 0.125  & 100\% \\
    & 0.6 & 0.128  & 100\% \\
    & 0.7 & 0.255  & 100\% \\
    & 0.8 & 0.380  & 100\% \\
    & 0.9 & 0.500  & 100\% \\
\midrule

\multirow{9}{*}{Shor $N=15$}
    & 0.1 & $-0.025$ & 30\%  \\
    & 0.2 & $-0.004$ & 56\%  \\
    & 0.3 & $-0.017$ & 40\%  \\
    & 0.4 & 0.034  & 42\%  \\
    & 0.5 & 0.030  & 52\%  \\
    & 0.6 & 0.112  & 98\%  \\
    & 0.7 & 0.200  & 100\% \\
    & 0.8 & 0.236  & 100\% \\
    & 0.9 & 0.298  & 100\% \\
\bottomrule
\end{tabular} 
\end{table}

\section{Related Work}
\label{sec:related}

Classical quantum-circuit simulation has been optimized along several
complementary axes. State-vector simulators reduce constant factors
through gate fusion, memory-aware layout, and parallelization.
Tensor-network methods exploit circuit structure, low entanglement,
or favorable contraction order; stabilizer and near-stabilizer methods
exploit Clifford structure; and distributed or GPU-accelerated tools
such as QuEST and Qiskit target larger circuits through hardware
parallelism~\cite{xu2023herculeantaskclassicalsimulation,weko_210570_1,jones_quest_2019,qiskicomm}.
These methods primarily optimize the representation and evolution of
the state. Our work instead isolates endpoint interference as the
object to be approximated.

The closest conceptual connection is to path-expansion and truncation
methods. Feynman's original path-integral viewpoint represents
amplitudes as sums over histories, and circuit path-integral
simulators similarly enumerate or sample histories before aggregating
endpoint amplitudes~\cite{FEYNPI,bernstein_quantum_nodate,feymanopt}.
Tensor-network contraction can also be read as a structured way of
reorganizing the same sum. Our ChAM model differs by making the
aggregation order itself nondeterministic and by terminating before
the endpoint sum is saturated.

Pauli propagation provides another relevant
comparison~\cite{Aharonov_2023,rudolph2025paulipropagationcomputationalframework,angrisani2025simulatingquantumcircuitsarbitrary,Dowling_2025,danna2025circuitcompression2dquantum}.
In Pauli-path methods, non-Clifford gates branch an observable into a
sum of Pauli terms, and truncation can be based on Pauli weight or
coefficient magnitude. Statistical interference sampling is
complementary: a Pauli-path approximation may discard low-coefficient Pauli branches generated by non-Clifford gates, reducing the number of terms propagated through the circuit. Statistical interference sampling can subsequently reduce the number of aggregation reactions performed among the surviving contributions at their common endpoints. The
Deutsch-Jozsa example illustrates this distinction: cancellation is
visible only after endpoint aggregation, precisely the step that
statistical interference sampling exposes and approximates. Pauli Path simulation of this circuit evolves as:

   \[
      \begin{split}
      \frac{1}{4} (I + Z_0 - Z_1 - Z_0Z_1) \rightarrow \\ 
      \frac{1}{4} (I + X_0 - X_1 - X_0X_1)  \rightarrow \\ 
       \frac{1}{4} (I - X_0 - X_1 + X_0X_1)  \rightarrow \\
           \frac{1}{4} (I - Z_0 - X_1 + Z_0X_1) 
      \end{split} 
    \]

The significant structural difference from our technique is that we follow many state paths and later combine amplitudes at common endpoints. Pauli-path simulation follows observable paths through the circuit. Because this Deutsch–Jozsa instance is entirely Clifford, each observable has only one Pauli path; providing no opportunity for truncation or approximation using frequency or amplitude based Pauli truncation. As the example above demonstrates, without gates such as rotations each path have equal coefficients, Pauli truncation does not present opportunities for fine grained truncation. Our work exposes a fine grained asymmetry in such cases by allowing coefficients to gradually build and cancel through interference, exposing a brand new plane for optimization.

Approximate simulation by neglecting small contributions also appears
in Monte Carlo methods, tensor-network truncation, and
decoherence-inspired
approximations~\cite{Fowler2012,Dennis2002,Joos1985,Zurek1991,Viola1998}.
The present work should be understood in that family. It does not
claim a new general-purpose simulator or a better worst-case
complexity bound; rather, it proposes a path-sum pruning primitive
that could be integrated into existing simulation frameworks whenever
endpoint interference can be exposed and scheduled explicitly.

\section{Conclusion and Future Work}
\label{sec:conc}

\paragraph*{Summary.}
We introduced statistical interference sampling, a path-sum
approximation in which endpoint interference is represented as a ChAM
reaction and may be stopped before saturation. The central idea is to
treat interference arithmetic as a separately schedulable resource
rather than an implicit consequence of matrix-vector multiplication.
The method remains exponential in the worst case, and its
accuracy depends on the endpoint-amplitude structure of the circuit.
The empirical results show that for circuits with strong amplitude
separation a significant fraction of endpoint reactions can
be omitted while preserving high output accuracy; for
broader distributions, such as the Shor instances studied here, the same threshold rule is less reliable but still
effective when combined with early termination. Early termination selection can be substantially more effective when the goal is instead to identify a task-valid endpoint,
particularly when the circuit produces strong amplitude separation. These objectives should not be conflated: on broader multi-output distributions, early termination
can achieve very high task success while remaining far from the exact Born distribution. The main conclusion is methodological: exposing interference as an explicit
aggregation step creates new opportunities for approximate simulation and for empirical studies of where partial path sums are already informative. The next steps are to derive error bounds for specific circuit families, replace scheduler-dependent nondeterminism with controlled sampling distributions, and integrate endpoint-interference pruning with tensor-network and Pauli-path simulators.

The instrumented executions provide direct evidence
for this mechanism: across the tested circuit families, the
transition to reliable early termination selection coincides
with the development of a positive separation between
the selected endpoint amplitude and the largest incorrect
ready-endpoint amplitude. This connects the empirical
behavior directly to the threshold-resolvable amplitude-gap characterization in Appendix \ref{sec:formal}.

\paragraph*{Physical interpretation.}
The construction is motivated by the path-integral language of
interfering histories, but it should not be read as a claim about how
quantum measurement or wavefunction collapse is realized physically.
The ChAM is a computational model for organizing and approximating a
path sum. Connecting the scheduling and thresholding rules to an
explicit open-system dynamics or decoherence mechanism would require a
separate analysis that we leave as an open question.

\begin{figure}[h]
\centering
\includegraphics[width=0.9\linewidth]{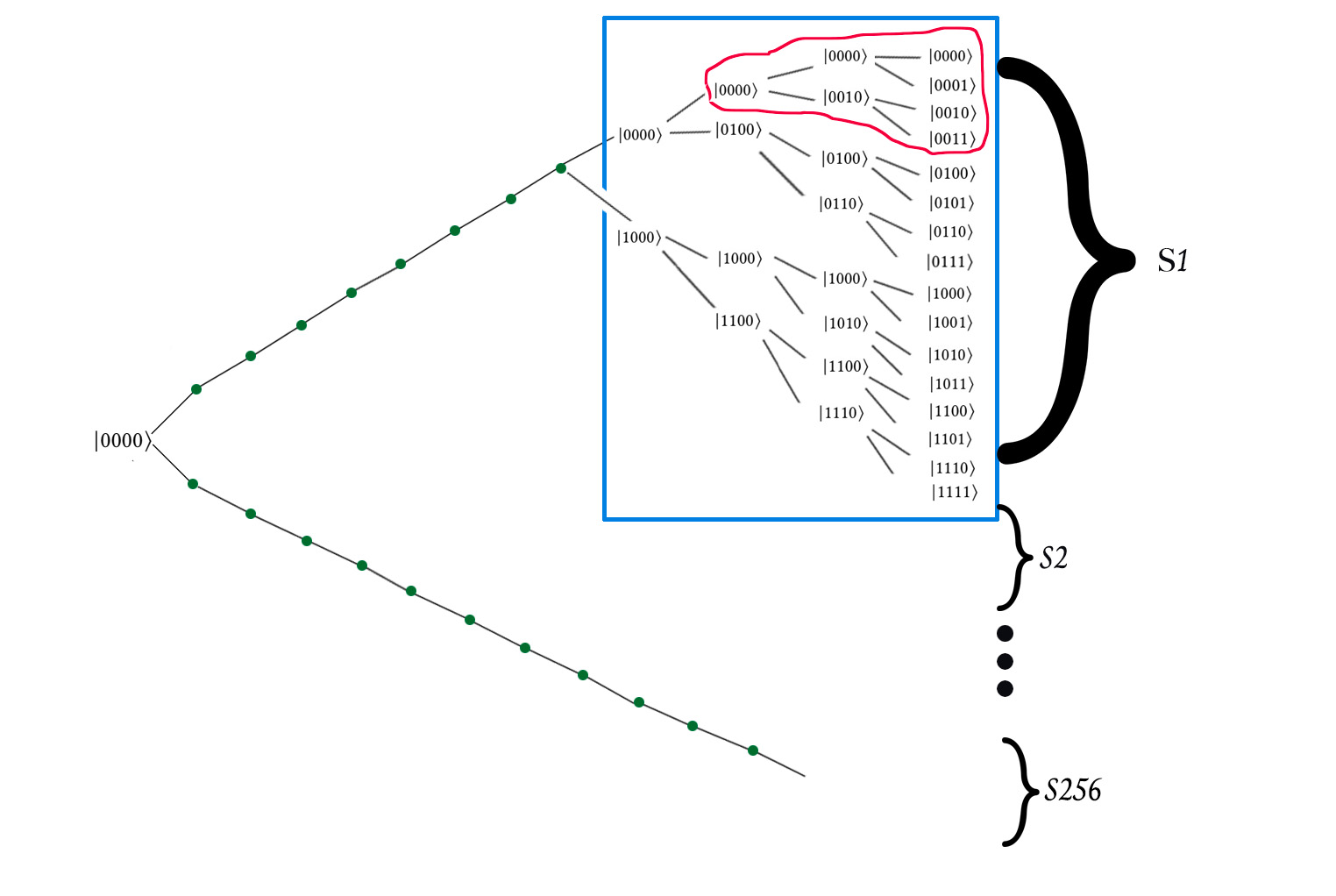}
\caption{Set patterns in the first iteration of Grover's algorithm
with 4 qubits. The blue rectangle marks a symmetric cut removing an
entire set of paths; the red circle marks an asymmetric cut removing
a subpath within one set.}
\label{fig:grovtree}
\end{figure}

\paragraph*{Exploiting symmetries.}
Our investigation into the concurrent simulation of Grover's algorithm
revealed hidden symmetries in the execution tree that suggest more
aggressive and structured truncations may be possible without
compromising accuracy. As illustrated in Figure~\ref{fig:abstractChAM},
the execution tree has a reflective structure: if one draws a
horizontal line through the root node, the set of states above the
line mirrors those below it, with interference occurring across this
axis between reflective pairs. Figure~\ref{fig:grovtree} visualizes
this for a single iteration of Grover's algorithm with a 4-qubit
tagged state search. Green nodes indicate Hadamard gates, while other
operations are omitted for clarity since they do not branch the tree.
The leaf nodes of this tree consist of 256 identical sets, each
containing 16 states, for a total of 4096 intermediate states. The
first of these sets is shown explicitly, and each set contributes one
correct molecule toward the final answer, forming the basis of the
symmetry. We conjecture that it may be possible to further optimize
simulations of Grover's algorithm by randomly cutting larger paths
within its execution tree, rather than simply truncating single nodes.

There are two ways to cut paths within this tree. The first, marked
by the red circle in Figure~\ref{fig:grovtree}, involves cutting a
subpath containing fewer than $n$ Hadamard nodes for an $n$-qubit
circuit. We refer to this as an \textit{asymmetric cut}, as it removes
some instances of certain states but not others. In Grover's algorithm
with the tagged state $\ket{0000}$, such a cut would remove an
intermediate state that contributes to the correct answer, reducing
the gap between the amplitudes of correct and incorrect solutions.
Nevertheless, due to the significant amplitude difference between
correct and incorrect outcomes and our method's reliance on early
sampling, Grover's algorithm can tolerate a certain number of such
cuts without substantially degrading the final result. In fact,
variance in thread progress during simulation already leads to similar
omissions, albeit less controlled. The precise level of tolerance will
depend on the specific circuit instance and tagged state. Importantly,
other algorithms such as Shor's algorithm may not tolerate asymmetric
cuts due to their stricter interference requirements.

The second class of cuts, marked by the blue rectangle in
Figure~\ref{fig:grovtree}, we term \textit{symmetric cuts}. Crucially,
these cuts are applied to the execution tree rather than to the
circuit itself: we are not removing entire layers of Hadamard gates,
but rather omitting the application of gates on entire subsets of
molecules within the simulation, which corresponds to allowing certain
threads to terminate early. Symmetric cuts differ from our existing
simulation methodology in two key ways. First, while our original
method randomly omits states at the leaf level independently and
without coordination across sets, a symmetric cut removes entire sets
at once, effectively eliminating one leaf from every possible state in
the tree, which leads to larger computational savings. Second, while
our current approach focuses on truncating single leaves, occasional
longer truncations from thread variance typically do not extend beyond
paths of length two in large circuits; symmetric cuts, by targeting
entire sets, are much deeper and promise greater efficiency gains.

The symmetry of these sets ensures that the impact of symmetric cuts
on the final measurement distribution is minimal. Any state $s \in
S_n$ carries an amplitude of the same magnitude as its counterpart in
another set $S_k$, such that if $\alpha_1\ket{s} \in S_n$ and
$\alpha_2\ket{s} \in S_k$, then $|\alpha_1| = |\alpha_2|$. For
Grover's algorithm on 4 qubits with the tagged state $\ket{0000}$,
the final amplitude of the correct outcome is $\alpha = -0.6875$,
while each incorrect outcome has amplitude $\beta = -0.1875$. Each
set $S_i$ contains a contribution of $\frac{-0.6875}{256}\ket{0000}$,
along with contributions $\pm\beta_i\ket{\cdots}$ from incorrect
states, where $\beta_i$ is much smaller than $-0.1875$. Removing a
set shifts the total amplitude of the correct state by
$\frac{-0.6875}{256}$ and affects the incorrect states collectively
by $15\beta_i$. After such a cut, amplitudes will no longer sum to
one and will require rescaling. Since all amplitudes are uniformly
reduced, rescaling preserves their proportions up to a negligible
error $\varepsilon$, retaining overall correctness of the final
distribution. Thus, symmetric cuts offer a principled way to achieve
deeper, more effective pruning of the execution tree, leveraging the
underlying symmetry of quantum circuits like Grover's to maintain
accuracy while greatly reducing computational effort. Exploring the
applicability of this strategy to other circuits, and characterizing
its theoretical error bounds, represents a promising avenue for future
research.

\paragraph*{Sampling and noise.}
We also see potential for extending this approach to algorithms that
sample from quantum probability distributions, such as Boson Sampling,
Fourier Sampling, and Q-Sampling, which are generally believed to be
classically intractable~\cite{bosonuniform,classicboson}. Modeling
interference under noise is another important challenge: fault-tolerant
algorithms achieve error correction by expanding the system and
increasing the number of qubits, which in turn enlarges the space of
possible superpositions and makes interference modeling more demanding.
Incorporating noise within the concurrent interference model is a
further direction we aim to investigate.

\section{Acknowledgment} 
This research was supported by the National Science Foundation under Grant No. CHE-2102610, including a Special Creativity Extension awarded to SSI.


\FloatBarrier
\appendix

\section{Formal Probabilistic Analysis}
\label{sec:formal}

Assuming an idealized arrival-order model introduced, we derive a bound on the probability
that a zero-contribution endpoint triggers early termination. Let
$N = 2^h$ be the number of terminal path contributions (leaves).
Each leaf $i$ contributes a basis vector
\[
  v_i = \alpha_i e_{x_i} \in \mathbb{C}^d
\]
where $\alpha_i$ is the complex amplitude, $x_i \in \{0,\ldots,d-1\}$
is the computational basis state, and $e_{x_i}$ is the corresponding
standard basis vector. The textbook final amplitude vector and
probability distribution are
\begin{equation}
    A = \sum^N_{i=1}v_i, \quad \|A\|^2 = 1, \qquad
    P^*(x) = |A_x|^2.
\end{equation}
Note that $A_x = 0$ for states in the zero-contribution set
$Z = \{x : A_x = 0\}$; these are endpoints at which all path
contributions cancel exactly.

Let $\pi$ be a uniform random permutation of $\{1,\ldots,N\}$,
denoting the arrival order of the leaves into the pool. After $k$
arrivals, the interfered amplitude vector is
\begin{equation}
  S(k) = \sum^k_{j=1}v_{\pi(j)}.
\end{equation}
The simulation halts at the first time the $\ell^\infty$ norm of
the partial sum exceeds the threshold $\threshold$:
\begin{equation}
    T = \min\bigl\{k : \|S(k)\|_{\infty} \geq \threshold \bigr\}.
\end{equation}
The output distribution is obtained by normalizing the partial sum
at halting:
\begin{equation}
    \hat{\psi} = \frac{S(T)}{\|S(T)\|_2}, \qquad
    Q_T(x) = \frac{|S_x(T)|^2}{\|S(T)\|^2_2}.
\end{equation}
Let $C$ be the set of correct or valid output states. The probability
of sampling an incorrect outcome is
\begin{equation}
    \Pr[\text{incorrect}] = \Pr[X \notin C], \quad X \sim Q_T,
\end{equation}
and the overall error relative to the textbook simulation is the
total variation distance
\begin{equation}
    \delta(Q_T, P^*) = \frac{1}{2} \sum_x |Q_T(x) - P^*(x)|.
    \label{eq:TV}
\end{equation}

Define the maximum leaf amplitude $b = \max_i\{|\alpha_i|\}$ and
let $N_x = |\{i : x_i = x\}|$ be the number of leaves with basis
state $x$. For a fixed $x \in Z$, the accumulated partial amplitude
after $k$ arrivals is
\[
  c_i = \alpha_{\pi(i)}\,
    \mathbb{I}[x_{\pi(i)} = x], \sum_{i=1}^N c_i = 0, |c_i| < b
\]
\begin{equation}
    S_x(k) = \sum_{j=1}^k c_j.
\end{equation}

\paragraph{Amplitude-gap characterization.}
We now quantify the amplitude-separation condition under
which threshold termination is reliable. Let
$I = \{0,\ldots,d-1\}\setminus C$ denote the non-empty set of
incorrect endpoints. Following the pairwise notion of
amplitude separation, define
\begin{equation}
  \Delta_{\mathrm{pair}}
  =
  \min_{\substack{c\in C\\i\in I}}
  |A_c-A_i|.
  \label{eq:pairwise-amplitude-gap}
\end{equation}
Because the stopping rule compares amplitude magnitudes,
rather than complex amplitudes themselves, the directly
relevant quantity is the threshold-resolvable gap
\begin{equation}
  \Delta_{\mathrm{th}}
  =
  a_C-a_I,
  \qquad
  a_C=\min_{c\in C}|A_c|,
  \qquad
  a_I=\max_{i\in I}|A_i|.
  \label{eq:threshold-amplitude-gap}
\end{equation}
A positive value $\Delta_{\mathrm{th}}>0$ means that every
correct endpoint has larger saturated amplitude magnitude
than every incorrect endpoint. When the correct and
incorrect amplitudes are phase-aligned,
$\Delta_{\mathrm{th}}=\Delta_{\mathrm{pair}}$. In general,
$\Delta_{\mathrm{pair}}>0$ alone is insufficient for
threshold separation, since amplitudes with different phases
may have equal magnitudes.

The relationship between the final amplitudes and the number
of processed leaves follows directly from the uniform random
arrival model. Each leaf appears among the first $k$ arrivals
with probability $k/N$, and hence
\begin{equation}
  \mathbb{E}[S_x(k)]
  =
  \frac{k}{N}A_x.
  \label{eq:expected-partial-amplitude}
\end{equation}
Equivalently, if
\[
  \overline{\alpha}_x
  =
  \frac{1}{N_x}
  \sum_{j:x_j=x}\alpha_j
  =
  \frac{A_x}{N_x}
\]
is the average leaf contribution to endpoint $x$, then the
expected number of endpoint-$x$ leaves required to reach
threshold $\mathcal T$ is approximately
\begin{equation}
  m_x(\mathcal T)
  \simeq
  \frac{\mathcal T}{|\overline{\alpha}_x|}
  =
  \frac{\mathcal T N_x}{|A_x|}.
  \label{eq:endpoint-leaves-to-threshold}
\end{equation}
Since an endpoint-$x$ leaf occurs with frequency $N_x/N$ in
the global arrival sequence, the corresponding expected
global stopping step is
\begin{equation}
  k_x(\mathcal T)
  \simeq
  \frac{N\mathcal T}{|A_x|}.
  \label{eq:global-steps-to-threshold}
\end{equation}
Thus correct endpoints reach a fixed threshold earlier than
incorrect endpoints whenever $\Delta_{\mathrm{th}}>0$. In
particular, the expected steps at which the weakest correct
endpoint and strongest incorrect endpoint reach the threshold
are
\begin{equation}
  k_C(\mathcal T)
  \simeq
  \frac{N\mathcal T}{a_C},
  \qquad
  k_I(\mathcal T)
  \simeq
  \frac{N\mathcal T}{a_I},
\end{equation}
and their expected separation is
\begin{equation}
  k_I(\mathcal T)-k_C(\mathcal T)
  \simeq
  \frac{N\mathcal T\Delta_{\mathrm{th}}}
       {a_Ca_I}.
  \label{eq:expected-crossing-separation}
\end{equation}
The separation therefore grows linearly with both the
threshold and the amplitude gap, although increasing the
threshold also causes a larger fraction of the leaves to be
processed.

To state a probabilistic guarantee, define the centered
partial-amplitude process
\begin{equation}
  D_x(k)
  =
  S_x(k)-\frac{k}{N}A_x.
\end{equation}
For $\epsilon>0$, consider the event
\begin{equation}
  \mathcal E_\epsilon
  =
  \left\{
    \max_x\max_{k\leq N}|D_x(k)|<\epsilon
  \right\}.
\end{equation}
The same finite-population martingale argument used below
for zero-contribution endpoints gives, up to the same
non-optimized constants,
\begin{equation}
  \Pr[\mathcal E_\epsilon^c]
  \leq
  \eta(\epsilon)
  :=
  \sum_{x=0}^{d-1}
  4\exp\left(
    -\frac{\epsilon^2}
    {4\left(N_xb^2+2b\epsilon/3\right)}
  \right).
  \label{eq:all-endpoint-deviation}
\end{equation}

Let $\rho\in(0,1]$ be the desired maximum fraction of
terminal leaves processed before termination. If
\begin{equation}
  \rho\Delta_{\mathrm{th}}>2\epsilon,
  \label{eq:gap-margin-condition}
\end{equation}
then the interval
\begin{equation}
  \rho a_I+\epsilon
  <
  \mathcal T
  \leq
  \rho a_C-\epsilon
  \label{eq:safe-threshold-interval}
\end{equation}
is nonempty. For every threshold in this interval, on the
event $\mathcal E_\epsilon$ no incorrect endpoint can cross
$\mathcal T$ during the first $\lceil\rho N\rceil$ arrivals,
while every correct endpoint has crossed it by arrival
$\lceil\rho N\rceil$. Consequently, the first endpoint to
cross the threshold is correct with probability at least
\begin{equation}
  \Pr[\text{correct early termination}]
  \geq
  1-\eta(\epsilon).
  \label{eq:correct-crossing-probability}
\end{equation}
This provides an explicit relationship among the threshold,
the amplitude gap, the amount of work performed, and the
expected accuracy of the early-selection rule.

For a threshold satisfying
Eq. \eqref{eq:safe-threshold-interval}, the event
$\mathcal E_\epsilon$ implies $T\leq\rho N$ and
\[
  |S_i(T)|
  \leq
  \rho|A_i|+\epsilon
  \qquad
  \text{for every } i\in I.
\]
Combining this bound with
$\|S(T)\|_2\geq\mathcal T$ at termination, and bounding the
complement of $\mathcal E_\epsilon$ by one, gives
\begin{equation}
  \Pr[X\notin C]
  \leq
  \min\!\left\{1,\,
    \eta(\epsilon)
    +
    \frac{
      \displaystyle
      \sum_{i\in I}
      \left(\rho|A_i|+\epsilon\right)^2
    }{\mathcal T^2}
  \right\}.
  \label{eq:gap-dependent-sampling-error}
\end{equation}
The first term bounds the probability that the partial sums
leave their concentration envelope; the second bounds the residual
incorrect probability mass at termination. Thus increasing
$\mathcal T$ reduces the normalized residual-error term, but
also increases the expected number of processed leaves
according to Eq. ~\eqref{eq:global-steps-to-threshold}.

\paragraph*{Bound on incorrect early termination.}
We first bound the probability that a zero-contribution endpoint
accumulates enough amplitude to trigger early termination. This
serves two purposes: it supports the claim that for a sufficiently
large threshold only correct states are likely to trigger a shutdown,
and it implicitly bounds the likelihood of unfavorable permutations
in which incorrect states undergo consecutive constructive
interference before being canceled.

Let $\mathcal{F}_k$ be the filtration generated by the first $k$ revealed leaves in a uniformly random permutation. The raw partial sum $S_x(k)$ is the reveal process associated with sampling without replacement. We apply a finite-population martingale concentration bound to the corresponding centered Doob process, separately to the real and imaginary parts, and then combine the two bounds by a union bound. In other terms, we construct a martingale from the random reveal process, then apply Freedman’s inequality to that martingale. Applying Freedman's inequality to the centered reveal martingale,
separately to the real and imaginary parts, gives a Bernstein-type
maximal bound of the form
\[
\Pr\!\left[
\max_{k\le N}|S_x(k)|\ge T
\right]
\le
4\exp\!\left(
-\frac{T^2}{4(N_x b^2+bT/3)}
\right),
\]
up to non-optimized constants.

Applying a union bound over all zero-contribution states,
\begin{align}
    &\Pr\!\left[\exists\, x \in Z : \max_{k}|S_x(k)|
    \geq \threshold\right] \nonumber \\
    &\quad \leq \sum_{x\in Z}4\exp\!\left(
    -\frac{\threshold^2}{4(N_x b^2 + b\threshold/3)}\right).
    \label{eq:zerobound}
\end{align}
This bound covers endpoints whose textbook amplitude is exactly
zero. We now extend it to account for endpoints that carry nonzero
but small amplitude in the full distribution.

\paragraph*{Full error bound.}
We establish a bound on the probability of sampling an incorrect
state at halting time $T$. Starting from the definition of $T$,
\begin{align}
    \|S(T)\|_{\infty} \geq \threshold
    &\Rightarrow \|S(T)\|_2 \geq \threshold \nonumber \\
    &\Rightarrow \Pr\!\left[X \notin C \mid S(T)\right]
    \leq \frac{\|S_{\mathrm{incorrect}}(T)\|_2^2}{\threshold^2}
    \label{eq:fullbound}
\end{align}
where $S_{\mathrm{incorrect}}(T)$ is the restriction of $S(T)$ to
incorrect states, defined by
\[
  S_{\mathrm{incorrect}}(T)_x =
  \begin{cases} S_x(T), & x \notin C \\ 0, & x \in C. \end{cases}
\]

For completeness, we derive this inequality explicitly. At time $T$
the pool corresponds to the unnormalized amplitude vector $S(T) \in
\mathbb{C}^d$. Normalizing and applying the Born rule,
\begin{align*}
  \Pr\!\left[X \notin C \mid S(T)\right]
  &= \sum_{x \notin C} \Pr\!\left[X = x \mid S(T)\right] \\
  &= \frac{\sum_{x\notin C}|S_x(T)|^2}{\|S(T)\|_2^2}
   = \frac{\|S_{\mathrm{incorrect}}(T)\|_2^2}{\|S(T)\|_2^2}.
\end{align*}
By the definition of $T$ and the norm inequality
$\|v\|_2 \geq \|v\|_\infty$ for all $v$,
\begin{align*}
  \|S(T)\|_2^2
  &= \sum_x |S_x(T)|^2
   \geq \max_x |S_x(T)|^2
   = \|S(T)\|^2_\infty.
\end{align*}
Therefore
\begin{align*}
  \|S(T)\|_2 &\geq \|S(T)\|_\infty \geq \threshold \\
  &\Rightarrow \|S(T)\|_2^2 \geq \threshold^2,
\end{align*}
which gives the stated bound
$\Pr[X \notin C \mid S(T)] \leq
\|S_{\mathrm{incorrect}}(T)\|_2^2 / \threshold^2$.

In summary, Eq. \ref{eq:fullbound} provides an upper bound for error in a single run that aligns with the empirical results. In our experiments, empirical error rates were consistently at least 15\% below this upper bound, suggesting that the bound is pessimistic in practice. We attribute this to the unfair scheduler discussed in Sec.~\ref{sec:scheduling}: because the formal analysis assumes a uniform random arrival order, the bound overestimates the probability that incorrect endpoints accumulate sufficient amplitude to trigger early termination.

This bounds remain loose due to the incomplete information about the OS and device level scheduler as discussed in \ref{sec:scheduling}. Mathematically we explain this as follows: Let $U$ denote the uniform distribution over leaf permutations assumed in this section, and let $Q_H$ denote the actual distribution induced by a particular execution environment $H$, including hardware, OS, runtime, scheduler policy, system load, task granularity. For the bad event

\[
  B_{\mathcal{T}} = \{ \exists x \notin C: \max\limits_{k}|S_x(k)| \geq \mathcal{T} \}
\]

one can write the generic scheduler-aware relation

\[
  Q_H(B_{ \mathcal{T}}) \leq U(B_{ \mathcal{T}}) +  d_{tv}(Q_H, U)
\]

where $d_{TV}$ is total variation distance between the actual and idealized arrival-order distributions. This equation identifies exactly what a scheduler-aware analysis would require: a bound on how far the actual permutation distribution is from uniform. The current algorithmic model contains no information from which that distance can be derived. Parallel-scheduling analyses generally require explicit assumptions about the task structure, scheduler policy, locality, memory behavior, and hardware execution model; results obtained for one combination do not automatically transfer to another. 

Estimating $Q_H$ is also not simply a matter of identifying the processor architecture. It depends on the Scala/JVM runtime, operating-system scheduler, work queues, preemption, cache state, competing system load, and heterogeneous-core assignments. The same hardware can therefore induce different arrival-order distributions across runtime configurations and executions.

A scheduler-aware probabilistic bound would require an explicit distribution $Q_H$ over completion orders for a specified hardware, operating system, runtime, scheduling policy, and workload. Such a distribution is not defined by the ChAM semantics and is not portable across execution environments. The semantics we profile and develop locally, may differ drastically on other machines due different systems level properties. Indeed, without restricting $Q_H$, a zero-contribution endpoint containing $m$ leaves of amplitude $+b$ and $m$ leaves of amplitude $-b$ can be made to cross any threshold $\mathcal{T}\leq mb$ by scheduling all positive leaves first. Hence no nontrivial scheduler-independent probabilistic bound is possible. Therefore we leave architecture-specific trace modeling or a controlled randomized scheduler as future work.

\section{Complementary Distribution-Distance Results}
\label{app:hellinger}

The main evaluation uses total-variation distance because it
is also the distributional error metric employed in the
formal analysis of Appendix \ref{sec:formal}. As a robustness check, we
additionally calculate Hellinger distance
\[
  H(Q_{\mathcal T},P^*)
  =
  \frac{1}{\sqrt{2}}
  \left[
    \sum_x
    \left(
      \sqrt{Q_{\mathcal T}(x)}
      -
      \sqrt{P^*(x)}
    \right)^2
  \right]^{1/2},
\]
and the associated quantity
\[
  F_H = 1-H^2.
\]
Table~\ref{tab:hellinger-results} reports these quantities for
representative normalized-Born configurations. The Hellinger
results lead to the same qualitative interpretation as TVD.
For example, Simon $n=3$ improves from
$H=0.559$ at $\mathcal{T}=0.1$ to $H=0.281$ at
$\mathcal{T}=0.9$, corresponding to an increase in
$F_H$ from $0.688$ to $0.921$. Shor $N=15$ remains
substantially farther from its saturated distribution,
although its Hellinger distance decreases from $0.804$ to
$0.605$ over the same threshold range.

\begin{table}[t]
\centering
\caption{
Complementary distribution-quality metrics for
normalized-Born thresholding. TVD is shown for comparison
with the main-text metric. $H$ is Hellinger distance and
$F_H=1-H^2$ is the corresponding Hellinger fidelity.}

\label{tab:hellinger-results}
\small
\setlength{\tabcolsep}{8pt}
\begin{tabular}{llrrr}
\toprule
Instance &
$\mathcal{T}$ &
TVD &
$H$ &
$F_H$ \\
\midrule

Tagged Grover
& 0.1 & 0.614 & 0.565 & 0.681 \\
& 0.3 & 0.384 & 0.392 & 0.847 \\
& 0.5 & 0.379 & 0.421 & 0.823 \\
& 0.7 & 0.409 & 0.369 & 0.864 \\
& 0.9 & 0.324 & 0.346 & 0.880 \\

\midrule

3-SAT Grover
& 0.1 & 0.696 & 0.688 & 0.526 \\
& 0.3 & 0.322 & 0.414 & 0.828 \\
& 0.5 & 0.314 & 0.416 & 0.827 \\
& 0.7 & 0.442 & 0.516 & 0.734 \\
& 0.9 & 0.309 & 0.407 & 0.835 \\

\midrule

Simon $n=3$
& 0.1 & 0.500 & 0.559 & 0.688 \\
& 0.3 & 0.400 & 0.479 & 0.771 \\
& 0.5 & 0.290 & 0.348 & 0.879 \\
& 0.7 & 0.240 & 0.366 & 0.866 \\
& 0.9 & 0.190 & 0.281 & 0.921 \\

\midrule

Simon $n=5$
& 0.1 & 0.533 & 0.673 & 0.547 \\
& 0.3 & 0.517 & 0.599 & 0.641 \\
& 0.5 & 0.498 & 0.590 & 0.652 \\
& 0.7 & 0.575 & 0.662 & 0.561 \\
& 0.9 & 0.435 & 0.536 & 0.713 \\

\midrule

Shor $N=15$
& 0.1 & 0.820 & 0.804 & 0.354 \\
& 0.3 & 0.780 & 0.819 & 0.329 \\
& 0.5 & 0.640 & 0.720 & 0.481 \\
& 0.7 & 0.720 & 0.702 & 0.508 \\
& 0.9 & 0.580 & 0.605 & 0.634 \\

\bottomrule
\end{tabular}
\end{table}

\section{Connection to Tensor-Network Contraction}
\label{app:tensor_network_connection}

In Sec. \ref{sec:related} we noted that tensor-network contraction may be viewed as a structured reorganization of the same coherent sum over contributions that is made explicit in the ChAM formulation. Here we make this connection precise. We show that statistical interference sampling can be applied whenever a tensor-network calculation is decomposed into additive complex contributions, and that this approximation acts on a different computational axis from conventional tensor-network bond truncation.

\subsection{A circuit path sum is a tensor-network contraction}

Consider the depth-$D$ circuit of Sec. \ref{sec:ChAM}
\[
U = U_D U_{D-1}\cdots U_1,
\]
with input basis state $\ket{x_0}$. Inserting a resolution of the identity in the computational basis between consecutive circuit layers gives
\begin{align}
A_D(x_D)
&=
\mel{x_D}{U_D U_{D-1}\cdots U_1}{x_0}
\nonumber\\
&=
\sum_{x_1,\ldots,x_{D-1}}
\prod_{d=1}^{D}
\mel{x_d}{U_d}{x_{d-1}}.
\label{eq:tn_path_sum}
\end{align}
This is exactly the path-sum expression used in Sec. \ref{sec:ChAM}. Each assignment
\[
\mu=(x_0,x_1,\ldots,x_D)
\]
defines one computational history with weight
\[
w(\mu)=
\prod_{d=1}^{D}
\mel{x_d}{U_d}{x_{d-1}},
\]
and Eq.~(\ref{eq:tn_path_sum}) is therefore
\[
A_D(x_D)=\sum_{\mu:x_D}w(\mu).
\]

The same expression is also a tensor-network contraction. Each matrix element of a circuit layer is a tensor, the intermediate basis labels $x_1,\ldots,x_{D-1}$ are contracted indices, and contracting an index means summing over all of its values. Thus, before any compression or change of representation, the circuit path sum and the corresponding tensor-network contraction are not two different physical descriptions: they are two ways of organizing the same scalar arithmetic.

More generally, for a tensor network with fixed external output $x$, let
$\boldsymbol{\gamma}$ denote an assignment of all internal contracted indices. Its contraction can always be written in the form
\begin{align}
A(x)
&=
\sum_{\boldsymbol{\gamma}}
w_{\boldsymbol{\gamma}}(x),
&
w_{\boldsymbol{\gamma}}(x)
&=
\prod_{v\in V}
T_v\!\left[\boldsymbol{\gamma},x\right],
\label{eq:general_tn_sum}
\end{align}
where $V$ is the set of tensors in the network. Equation (\ref{eq:general_tn_sum}) has exactly the additive structure required by the ChAM interference rule.

There is one useful distinction. In the direct circuit tensor network, an assignment of the intermediate computational-basis indices corresponds directly to a circuit history $\mu$. After tensors have been contracted, factorized, or compressed, an internal-index assignment need no longer correspond to one unique physical circuit history; it may represent a regrouping of many such histories. The equality of the saturated sums is unaffected by this regrouping.

\subsection{MPS as a simple specialization}

An MPS makes the same point in a familiar one-dimensional notation. For $n$ qubits, write
\begin{align}
\psi(x_1,\ldots,x_n)
=
\sum_{\alpha_1,\ldots,\alpha_{n-1}}
\prod_{j=1}^{n}
\phi^{[j]}_{\alpha_{j-1},\alpha_j}(x_j),
\label{eq:mps_basic}
\end{align}
where $\alpha_0=\alpha_n=1$ and the $\alpha_j$ are internal, or bond, indices. Defining
\[
w_{\boldsymbol{\alpha}}(\bar{x})
=
\prod_{j=1}^{n}
\phi^{[j]}_{\alpha_{j-1},\alpha_j}(x_j),
\]
Eq.~(\ref{eq:mps_basic}) becomes simply
\begin{equation}
\psi(\bar{x})
=
\sum_{\boldsymbol{\alpha}}
w_{\boldsymbol{\alpha}}(\bar{x}).
\label{eq:mps_sum_terms}
\end{equation}

The important point is that
$w_{\boldsymbol{\alpha}}(\bar{x})$ is a complex
\emph{amplitude contribution}, not a probability. The probability of observing $\bar{x}$ is obtained only after the contributions have been coherently combined:
\begin{align}
P(\bar{x})
&=
\left|
\sum_{\boldsymbol{\alpha}}
w_{\boldsymbol{\alpha}}(\bar{x})
\right|^2
\nonumber\\
&=
\sum_{\boldsymbol{\alpha}}
\left|w_{\boldsymbol{\alpha}}(\bar{x})\right|^2
+
\sum_{\boldsymbol{\alpha}\neq\boldsymbol{\beta}}
w_{\boldsymbol{\alpha}}(\bar{x})
w_{\boldsymbol{\beta}}(\bar{x})^* .
\label{eq:mps_interference}
\end{align}
The second term contains the constructive and destructive cross-terms. Thus the interference exposed by the ChAM does not disappear in an MPS representation; it is contained in the summation over the contracted internal indices.

It is useful  to distinguish the coherent aggregation operation from the interference terms that it produces. The contraction 

\[ \psi(\bar{x}) = \sum_{\boldsymbol{\alpha}} w_{\boldsymbol{\alpha}}(\bar{x}) \] 

is an amplitude-level coherent sum over internal-index assignments. Interference becomes explicit when the squared magnitude of this sum is formed. For a fixed MPS representation, define the net cross-term contribution at endpoint $\bar{x}$ as 

\begin{align} I_{\mathrm{MPS}}(\bar{x}) &= \sum_{\boldsymbol{\alpha}\neq\boldsymbol{\beta}} w_{\boldsymbol{\alpha}}(\bar{x}) w_{\boldsymbol{\beta}}(\bar{x})^* \nonumber\\ &= P(\bar{x}) - \sum_{\boldsymbol{\alpha}} \left|w_{\boldsymbol{\alpha}}(\bar{x})\right|^2 . \label{eq:mps_cross_interference} \end{align} 

The quantity $P(\bar{x})$ is a probability and hence positive-definite and real. The sum of diagonal elements is real, given that its a sum of absolute values which makes the sum of off-diagonal contributions, $I_{\mathrm{MPS}}(\bar{x})$ also real.
A positive value denotes net constructive interference for this decomposition, while a negative value denotes net destructive interference. Thus, in an MPS, contraction over the internal indices is the arithmetic operation through which amplitude contributions are coherently aggregated, while the off-diagonal cross-terms in Eq.~(\ref{eq:mps_cross_interference}) are the corresponding interference contributions exposed by the Born rule.

\[ \underbrace{ \sum_{\boldsymbol{\alpha}} w_{\boldsymbol{\alpha}}(\bar{x}) }_{\text{coherent amplitude aggregation}} \quad\Longrightarrow\quad \underbrace{ \sum_{\boldsymbol{\alpha}\neq\boldsymbol{\beta}} w_{\boldsymbol{\alpha}}(\bar{x}) w_{\boldsymbol{\beta}}(\bar{x})^* }_{\text{interference contribution}} \]

This gives a direct formal correspondence with the ChAM
interference reaction. For a fixed endpoint $\bar{x}$,
associate an additive tensor-network contribution with the
molecule
\[
w_{\boldsymbol{\alpha}}(\bar{x})\ket{\bar{x}}.
\]
Then the ChAM reaction
\[
a_1\ket{\bar{x}},a_2\ket{\bar{x}}
\longrightarrow
(a_1+a_2)\ket{\bar{x}}
\]
is precisely a pairwise reduction of terms in
Eq.~(\ref{eq:mps_sum_terms}). The interference effect of such
a reduction is made explicit by
\begin{align}
\left|a_1+a_2\right|^2
&=
|a_1|^2+|a_2|^2
+
2\operatorname{Re}\!\left(a_1a_2^*\right).
\label{eq:pairwise_interference}
\end{align}
The final term is the pairwise interference contribution:
it is positive when the two amplitudes reinforce and negative
when they cancel. The ChAM reaction therefore provides an explicit,
interruptible realization of an additive aggregation that
appears implicitly within the MPS contraction: each ChAM interference reaction
performs one step of the same coherent amplitude aggregation
implemented by contraction of the MPS internal indices.

If all such reductions are allowed to saturate, repeated
pairwise aggregation yields
\[
\left(
\sum_{\boldsymbol{\alpha}}
w_{\boldsymbol{\alpha}}(\bar{x})
\right)\ket{\bar{x}}
=
\psi(\bar{x})\ket{\bar{x}},
\]
so the saturated ChAM aggregation and the saturated
tensor-network contraction return the same endpoint
amplitude. Applying the Born rule to that amplitude then
generates exactly the complete set of diagonal and
off-diagonal terms in Eq.~(\ref{eq:mps_interference}).

This correspondence is algebraic rather than a claim that the virtual MPS indices themselves are observable trajectories. In particular, equivalent MPS representations may use different internal gauges, and a gauge transformation can redistribute the individual contributions $w_{\boldsymbol{\alpha}}(\bar{x})$ without changing their fully contracted sum. Consequently, neither the individual $w_{\boldsymbol{\alpha}}(\bar{x})$ nor the decomposition of $P(\bar{x})$ into its diagonal part and $I_{\mathrm{MPS}}(\bar{x})$ should be interpreted as representation-independent physical quantities. What is representation independent is the fully contracted amplitude $\psi(\bar{x})$ and hence the observable probability $|\psi(\bar{x})|^2$. The ChAM--MPS correspondence is therefore a correspondence between additive complex reductions. This shared additive structure makes the ChAM interference reaction applicable to an MPS contraction.

\subsection{Applying statistical interference to a tensor-network contraction}

Equation~(\ref{eq:general_tn_sum}) is sufficient to apply the stopping rule of this paper, but enumerating every virtual-index assignment individually would generally destroy the computational advantage of a tensor-network representation. A more useful implementation is obtained from \emph{slicing}.

Suppose a set of internal tensor-network indices is selected for slicing. Fixing those indices produces independent sub-contractions indexed by $s\in\mathcal{S}$. Exact contraction then has the additive form
\begin{equation}
A(x)
=
\sum_{s\in\mathcal{S}} A_s(x),
\label{eq:tn_slices}
\end{equation}
where each $A_s(x)$ is obtained by contracting the remaining unsliced network. A slice may therefore summarize many individual circuit histories while remaining an additive complex contribution to the final endpoint amplitude.

If the slices are processed in an order $\pi$, define
\begin{equation}
S_x(k)
=
\sum_{j=1}^{k} A_{\pi(j)}(x).
\label{eq:tn_partial_sum}
\end{equation}
The statistical-interference stopping rule may then be applied without modification:
\begin{equation}
\tau_T
=
\min\left\{
k:
\max_x |S_x(k)|\geq T
\right\}.
\label{eq:tn_stop}
\end{equation}
At $\tau_T$, the available endpoint amplitudes may be normalized and sampled as in Sec. \ref{sec:threshold}, or the first endpoint crossing $T$ may be returned under the early termination policy.

Hence, at the level of additive arithmetic, the molecules need not be individual circuit leaves. They may instead be tensor-network slices or other independently contractible blocks. A saturated execution sums every block and recovers the ordinary tensor-network result; statistical interference sampling asks whether the sum can be stopped before all of those blocks have been accumulated.

The probabilistic bounds in Appendix \ref{sec:formal} depend on assumptions about the order and magnitude of the arriving contributions. They therefore do not automatically apply to an arbitrary deterministic tensor-network contraction order. However, if the additive blocks in Eq. ~(\ref{eq:tn_slices}) are processed under the same random-order model assumed in Appendix~A, the same partial-sum argument applies, with the tensor-network blocks replacing the terminal path contributions. Different slicing or scheduling strategies will in general produce different partial-sum statistics and therefore require corresponding bounds.

\subsection{Complementarity with bond truncation}

The distinction between tensor-network truncation and statistical interference sampling is particularly important. Standard MPS methods reduce computational cost by limiting a bond dimension, commonly by discarding small singular values during an SVD-based compression. This changes the state that is subsequently propagated or contracted. Statistical interference sampling instead changes how completely the surviving additive contributions are accumulated.

The two approximations can therefore be composed. Let $A$ denote the exact saturated amplitude vector and let $\widetilde{A}$ denote the amplitude vector obtained after tensor-network truncation. Suppose the tensor-network approximation satisfies
\begin{equation}
\left\|A-\widetilde{A}\right\|_2
\leq
\varepsilon_{\mathrm{TN}}.
\label{eq:tn_error}
\end{equation}
For the standard sequential SVD construction, $\varepsilon_{\mathrm{TN}}$ can, for example, be bounded using the discarded singular values of the truncated bonds. The precise value of the bound is not important for the argument below; only Eq. ~(\ref{eq:tn_error}) is required.

Recall from Appendix \ref{sec:formal} the saturated threshold-resolvable gap
\begin{equation}
\Delta_{\mathrm{th}}
=
a_C-a_I,
\qquad
a_C=\min_{c\in C}|A_c|,
\qquad
a_I=\max_{i\notin C}|A_i|.
\label{eq:original_gap_tn}
\end{equation}
Since every component satisfies
\[
|\widetilde{A}_x-A_x|
\leq
\left\|\widetilde{A}-A\right\|_2
\leq
\varepsilon_{\mathrm{TN}},
\]
the reverse triangle inequality gives
\begin{align}
\min_{c\in C}|\widetilde{A}_c|
&\geq
a_C-\varepsilon_{\mathrm{TN}},
\\
\max_{i\notin C}|\widetilde{A}_i|
&\leq
a_I+\varepsilon_{\mathrm{TN}}.
\end{align}
Consequently the threshold-resolvable gap after tensor-network truncation obeys
\begin{equation}
\widetilde{\Delta}_{\mathrm{th}}
\geq
\Delta_{\mathrm{th}}
-
2\varepsilon_{\mathrm{TN}}.
\label{eq:gap_after_tn}
\end{equation}

Equation(\ref{eq:gap_after_tn}) gives a simple sufficient condition for the two approximations to remain compatible:
\begin{equation}
\Delta_{\mathrm{th}}
>
2\varepsilon_{\mathrm{TN}}.
\label{eq:tn_preserve_gap}
\end{equation}
Under this condition, tensor-network truncation is guaranteed not to eliminate the positive saturated amplitude separation used by the early termination analysis of Appendix \ref{sec:formal}.

The connection can be made more explicit. Let $\rho\in(0,1]$ denote the desired maximum fraction of additive tensor-network blocks processed, and suppose that their random arrival order obeys the same concentration event as Appendix \ref{sec:formal},
\begin{equation}
\max_x\max_{k}
\left|
S_x(k)-\frac{k}{|\mathcal{S}|}\widetilde{A}_x
\right|
<
\varepsilon_{\mathrm{ord}}.
\label{eq:tn_concentration}
\end{equation}
A sufficient condition for a threshold separating correct and incorrect endpoints before a fraction $\rho$ of the blocks has been processed is then
\begin{equation}
\rho
\left(
\Delta_{\mathrm{th}}
-
2\varepsilon_{\mathrm{TN}}
\right)
>
2\varepsilon_{\mathrm{ord}}.
\label{eq:combined_condition}
\end{equation}
Indeed, whenever Eq.~(\ref{eq:combined_condition}) holds, the interval
\begin{equation}
\rho
\left(a_I+\varepsilon_{\mathrm{TN}}\right)
+
\varepsilon_{\mathrm{ord}}
<
T
\leq
\rho
\left(a_C-\varepsilon_{\mathrm{TN}}\right)
-
\varepsilon_{\mathrm{ord}}
\label{eq:combined_threshold}
\end{equation}
is nonempty. On the concentration event
Eq. (\ref{eq:tn_concentration}), no incorrect endpoint can reach such a threshold during the first
$\rho|\mathcal{S}|$ contributions, whereas the correct endpoints reach it by that boundary. Thus the same early termination guarantee derived in Appendix \ref{sec:formal} continues to hold, with an effective gap reduced by at most
$2\varepsilon_{\mathrm{TN}}$.

This gives a direct interpretation of complementarity. Tensor-network compression consumes part of the available amplitude-separation margin through
$\varepsilon_{\mathrm{TN}}$, while statistical interference sampling consumes another part through incomplete aggregation and its fluctuation
$\varepsilon_{\mathrm{ord}}$. If the original circuit has a sufficiently large $\Delta_{\mathrm{th}}$, both approximations may be applied while retaining a positive threshold-resolvable margin.

\bibliographystyle{quantum}
\bibliography{cites,wavepacket}

@mastersthesis{feymanopt,
  author = {Ferreira, David Alves Campos},
  title = {{Feynman} path-sum quantum computer simulator},
  school = {Universidade do Minho},
  year = {2023},
  nolink = {},
}

@book{FEYNPI,
  author = {Feynman, Richard P. and Hibbs, Albert R.},
  title = {{Quantum} Mechanics and Path Integrals},
  publisher = {McGraw-Hill},
  address = {New York},
  year = {1965},
  nolink = {},
}

@inproceedings{classicboson,
  author = {Clifford, Peter and Clifford, Rapha\"{e}l},
  title = {The Classical Complexity of {Boson} Sampling},
  booktitle = {Proceedings of the Twenty-Ninth Annual ACM-SIAM Symposium on Discrete Algorithms},
  series = {SODA '18},
  pages = {146--155},
  publisher = {Society for Industrial and Applied Mathematics},
  address = {USA},
  location = {New Orleans, Louisiana},
  year = {2018},
  nolink = {},
}

@article{bosonuniform,
  author = {Aaronson, Scott and Arkhipov, Alex},
  title = {{BosonSampling} is Far from Uniform},
  journal = {Quantum Info. Comput.},
  volume = {14},
  number = {15--16},
  pages = {1383--1423},
  publisher = {Rinton Press, Incorporated},
  address = {Paramus, NJ},
  year = {2014},
  month = {nov},
  nolink = {},
}

@book{NielsenChuang2010,
  author = {Nielsen, Michael A. and Chuang, Isaac L.},
  title = {{Quantum} Computation and {Quantum} Information},
  publisher = {Cambridge University Press},
  address = {Cambridge},
  edition = {10th anniversary ed.},
  year = {2010},
  nolink = {},
}

@article{doi:10.1137/S0097539796298637,
  author = {Simon, Daniel R.},
  title = {On the Power of {Quantum} Computation},
  journal = {SIAM Journal on Computing},
  volume = {26},
  number = {5},
  pages = {1474--1483},
  year = {1997},
  doi = {10.1137/S0097539796298637},
}

@article{doi:10.1137/S0036144598347011,
  author = {Shor, Peter W.},
  title = {Polynomial-Time Algorithms for Prime Factorization and Discrete Logarithms on a {Quantum} Computer},
  journal = {SIAM Review},
  volume = {41},
  number = {2},
  pages = {303--332},
  year = {1999},
  doi = {10.1137/S0036144598347011},
}

@article{deutschJozsa,
  author = {Deutsch, David and Jozsa, Richard},
  title = {Rapid Solution of Problems by {Quantum} Computation},
  journal = {Proc. R. Soc. Lond. A},
  volume = {439},
  number = {1907},
  pages = {553--558},
  year = {1992},
  doi = {10.1098/rspa.1992.0167},
}

@inproceedings{bernstein_quantum_nodate,
  author = {Bernstein, Ethan and Vazirani, Umesh},
  title = {{Quantum} Complexity Theory},
  booktitle = {Proceedings of the Twenty-Fifth Annual ACM Symposium on Theory of Computing},
  series = {STOC '93},
  pages = {11--20},
  publisher = {Association for Computing Machinery},
  address = {New York, NY, USA},
  location = {San Diego, California, USA},
  year = {1993},
  doi = {10.1145/167088.167097},
}

@techreport{weko_210570_1,
  author = {Horii, Hiroshi and Doi, Jun},
  title = {Optimization of {Quantum} Computing Simulation with Gate Fusion},
  number = {23},
  institution = {IBM Quantum, IBM Research Tokyo},
  year = {2021},
  month = {mar},
  nolink = {},
}

@article{jones_quest_2019,
  author = {Jones, Tyson and Brown, Anna and Bush, Ian and Benjamin, Simon C.},
  title = {{QuEST} and {High} {Performance} {Simulation} of {Quantum} {Computers}},
  journal = {Scientific Reports},
  volume = {9},
  number = {1},
  pages = {10736},
  year = {2019},
  month = {jul},
  doi = {10.1038/s41598-019-47174-9},
  url = {https://www.nature.com/articles/s41598-019-47174-9},
}

@misc{qiskicomm,
  author = {{Qiskit Development Team}},
  title = {{Qiskit} Transpiler Pass: {CommutativeCancellation}},
  year = {2024},
  url = {https://docs.quantum.ibm.com/api/qiskit/qiskit.transpiler.passes.CommutativeCancellation},
}

@article{Joos1985,
  author = {E. Joos and H. D. Zeh},
  title = {The emergence of classical properties through interaction with the environment},
  journal = {Zeitschrift für Physik B},
  volume = {59},
  number = {2},
  pages = {223--243},
  year = {1985},
  doi = {10.1007/BF01725541},
}

@article{Zurek1991,
  author = {Wojciech H. Zurek},
  title = {Decoherence and the transition from quantum to classical},
  journal = {Physics Today},
  volume = {44},
  pages = {36--44},
  year = {1991},
  doi = {10.1063/1.881293},
}

@article{Viola1998,
  author = {Lorenza Viola and Seth Lloyd},
  title = {Dynamical suppression of decoherence in two-state quantum systems},
  journal = {Phys. Rev. A},
  volume = {58},
  number = {4},
  pages = {2733--2744},
  year = {1998},
  doi = {10.1103/PhysRevA.58.2733},
}

@article{Fowler2012,
  author = {Austin G. Fowler and Matteo Mariantoni and John M. Martinis and Andrew N. Cleland},
  title = {Surface codes: Towards practical large-scale quantum computation},
  journal = {Phys. Rev. A},
  volume = {86},
  number = {3},
  pages = {032324},
  year = {2012},
  doi = {10.1103/PhysRevA.86.032324},
}

@article{Dennis2002,
  author = {Eric Dennis and Alexei Kitaev and Andrew Landahl and John Preskill},
  title = {Topological quantum memory},
  journal = {Journal of Mathematical Physics},
  volume = {43},
  number = {9},
  pages = {4452--4505},
  year = {2002},
  doi = {10.1063/1.1499754},
}

@article{Deutsch1985,
  author = {David Deutsch},
  title = {{Quantum} Theory, the Church-Turing Principle and the Universal {Quantum} Computer},
  journal = {Proc. R. Soc. Lond. A},
  volume = {400},
  pages = {97--117},
  year = {1985},
  doi = {10.1098/rspa.1985.0070},
}

@article{Berry1992,
  author = {Berry, G\'{e}rard and Boudol, G\'{e}rard},
  title = {The {Chemical Abstract Machine}},
  journal = {Theor. Comput. Sci.},
  volume = {96},
  number = {1},
  pages = {217--248},
  publisher = {Elsevier Science Publishers Ltd.},
  address = {GBR},
  year = {1992},
  month = {apr},
  doi = {10.1016/0304-3975(92)90185-I},
}

@inproceedings{10.1145/237814.237866,
  author = {Grover, Lov K.},
  title = {A Fast {Quantum} Mechanical Algorithm for Database Search},
  booktitle = {Proceedings of the Twenty-Eighth Annual ACM Symposium on Theory of Computing},
  series = {STOC '96},
  pages = {212--219},
  publisher = {Association for Computing Machinery},
  address = {New York, NY, USA},
  location = {Philadelphia, Pennsylvania, USA},
  year = {1996},
  doi = {10.1145/237814.237866},
}

@misc{xu2023herculeantaskclassicalsimulation,
  author = {Xiaosi Xu and Simon Benjamin and Jinzhao Sun and Xiao Yuan and Pan Zhang},
  title = {A Herculean task: Classical simulation of quantum computers},
  year = {2023},
  url = {https://arxiv.org/abs/2302.08880},
  eprint = {2302.08880},
  archivePrefix = {arXiv},
  primaryClass = {quant-ph},
}

@inproceedings{Aharonov_2023,
  author = {Aharonov, Dorit and Gao, Xun and Landau, Zeph and Liu, Yunchao and Vazirani, Umesh},
  title = {A Polynomial-Time Classical Algorithm for Noisy Random Circuit Sampling},
  booktitle = {Proceedings of the 55th Annual ACM Symposium on Theory of Computing},
  series = {STOC '23},
  pages = {945--957},
  publisher = {ACM},
  year = {2023},
  month = {jun},
  doi = {10.1145/3564246.3585234},
  collection = {STOC '23},
}

@misc{rudolph2025paulipropagationcomputationalframework,
  author = {Manuel S. Rudolph and Tyson Jones and Yanting Teng and Armando Angrisani and Zoë Holmes},
  title = {{Pauli} Propagation: A Computational Framework for Simulating {Quantum} Systems},
  year = {2025},
  url = {https://arxiv.org/abs/2505.21606},
  eprint = {2505.21606},
  archivePrefix = {arXiv},
  primaryClass = {quant-ph},
}

@misc{angrisani2025simulatingquantumcircuitsarbitrary,
  author = {Armando Angrisani and Antonio A. Mele and Manuel S. Rudolph and M. Cerezo and Zoë Holmes},
  title = {Simulating quantum circuits with arbitrary local noise using {Pauli} Propagation},
  year = {2025},
  url = {https://arxiv.org/abs/2501.13101},
  eprint = {2501.13101},
  archivePrefix = {arXiv},
  primaryClass = {quant-ph},
}

@article{Dowling_2025,
  author = {Dowling, Neil and Kos, Pavel and Turkeshi, Xhek},
  title = {Magic Resources of the {Heisenberg} Picture},
  journal = {Physical Review Letters},
  volume = {135},
  number = {5},
  pages = {050401},
  publisher = {American Physical Society (APS)},
  year = {2025},
  month = {jul},
  doi = {10.1103/p7xt-s9nz},
}

@misc{danna2025circuitcompression2dquantum,
  author = {Matteo D'Anna and Yuxuan Zhang and Roeland Wiersema and Manuel S. Rudolph and Juan Carrasquilla},
  title = {Circuit compression for 2D quantum dynamics},
  year = {2025},
  url = {https://arxiv.org/abs/2507.01883},
  eprint = {2507.01883},
  archivePrefix = {arXiv},
  primaryClass = {quant-ph},
}
\end{document}